\documentclass[journal]{IEEEtran}
\IEEEoverridecommandlockouts
\usepackage{cite}
\usepackage{amsmath,amssymb,amsfonts}
\usepackage{graphicx}
\usepackage{textcomp}
\usepackage{xcolor}
\usepackage{multirow}
\usepackage{threeparttable}
\usepackage{url}
\usepackage{siunitx}
\usepackage{soul}
\usepackage{booktabs}
\usepackage{color,colortbl}
\usepackage[linesnumbered,ruled,vlined]{algorithm2e}
\usepackage{algpseudocode}
\usepackage[switch]{lineno}
\pdfoutput=1

\definecolor{bgc}{HTML}{24292E}
\begin{document}

\hyphenation{op-tical net-works semi-conduc-tor}

\title{
	Neonatal Bowel Sound Detection Using 
	Convolutional Neural Network and Laplace Hidden Semi-Markov Model
}
\author{Chiranjibi Sitaula, Jinyuan He, Archana Priyadarshi, Mark Tracy, Omid Kavehei, Murray Hinder, Anusha Withana, Alistair McEwan, Faezeh Marzbanrad,~\IEEEmembership{Senior Member,~IEEE}
\thanks{This work was supported by the Monash Data Future Institute (MDFI).}
\thanks{C. Sitaula, J. He and F. Marzbanrad are with the Department of Electrical and Computer Systems Engineering, Monash University, Clayton, VIC 3800, Australia.}
\thanks{A. Priyadarshi, M. Tracy and M. Hinder are with Department of Paediatrics and Child Health at The University of Sydney and Westmead Hospital}
\thanks{O. Kavehei and A. McEwan are with School of Biomedical Engineering, The University of Sydney.}
\thanks{A. Withana is with the School of Computer Science, The University of Sydney.}
\thanks{email: faezeh.marzbanrad@monash.edu}
}

\maketitle

\begin{abstract}
Abdominal auscultation is a convenient, safe and inexpensive method to assess bowel conditions, which is essential in neonatal care. It helps early detection of neonatal bowel dysfunctions and allows timely intervention. This paper presents a neonatal bowel sound detection method to assist the auscultation. Specifically, 
a Convolutional Neural Network (CNN) is proposed to classify peristalsis and non-peristalsis sounds. The classification is then optimized using a Laplace Hidden Semi-Markov Model (HSMM). 
The proposed method is validated on abdominal sounds from 49 newborn infants admitted to our tertiary Neonatal Intensive Care Unit (NICU). The results show that the method can effectively detect bowel sounds with accuracy and area  under  curve  (AUC) score being 89.81\% and 83.96\% respectively, outperforming 13 baseline methods.
Furthermore, the proposed Laplace HSMM refinement strategy is proven capable to enhance other bowel sound detection models. 
The outcomes of this work have the potential to facilitate future telehealth applications for neonatal care.
The source code of our work can be found at: \url{https://bitbucket.org/chirudeakin/neonatal-bowel-sound-classification/.}

\end{abstract}


\section{Introduction}
Bowel sound, i.e., peristalsis sound, denotes the gurgling, rumbling, or growling noises from abdomen, which is produced by the movement of food, gas and fluids during intestinal peristalsis \cite{liu2018bowel}. 
It contains valuable information of gastrointestinal condition. Bowel motility disorder can be reflected as an abnormality or lack of peristalsis sound.
Auscultation of bowel sound is therefore a routine assessment in neonatal care \cite{queensland2019routine}, which is particularly important to newborns who are admitted to NICU. The assessment enables early diagnosis of bowel dysfunctions of newborns, such as feeding intolerance and intestinal obstruction \cite{dumas2013feasibility,hill2008stethoscope}, and Necrotising Enterocolitis (NEC), a potentially fatal disease affecting 6\% of infants born with under 1500 grams birth weight \cite{calvert2020necrotising}. Extremely preterm newborns have immature gut and frequent dysmotility shown clinically by feed intolerance bilious gastric aspirates and abdominal distension. Most of these infants have recovering respiratory distress syndrome needing nasal Continuous Positive Airway Pressure (CPAP). Nasal CPAP commonly distends the gut with gas into the stomach leading to ``CPAP belly syndrome" \cite{priyadarshi2020continuous}. This is frequently associated with increased dysmotility and clinical confusion with early NEC. Alteration of bowel sounds is a part of assessment for ealry NEC development.  

Traditionally, neonatal bowel condition has been qualitatively assessed by placing a standard acoustic stethoscope on the abdominal wall of the neonate. By listening to the abdominal sounds in all four quadrants (right upper, right lower, left upper and left lower quadrants), the bowel condition can be roughly determined based on whether the auscultated bowel sound is frequent, diminished, or absent. 
Although it is simple and easy to perform, the assessment relies heavily on the skill and knowledge of the on-site specialist and the application of the stethoscope is limited by the varying quality and irreproducibility of the collected bowel sounds. 
With the development of digital stethoscope (DS), a more dynamic and objective way can be provided to monitor neonatal bowel conditions. In contrast to the acoustic stethoscope, digital stethoscope converts acoustic sounds to electrical signals and allows storage and transmission of the signals for computerized process, medical crowdsourcing, and telehealth applications.  
However, despite the improvements in auscultation introduced by digital stethoscope, interpretation of neonatal bowel sounds still imposes a heavy demand on workload and skilled personnel \cite{ramanathan2019digital}. 
Automatically detecting and locating the bowel sounds in digital stethoscope recordings would enable a more objective, effective, and efficient diagnosis \cite{yin2018bowel}. 

Automatic detection of bowel sound has been investigated in previous studies. 
A \textit{Multi-layer Perceptron} (MLP)-based method for long-term abdominal sound monitoring was presented in \cite{dimoulas2008bowel}. The study explored the rationale of using time-frequency features and wavelet-adapted parameters for bowel sound recognition. However, the method of evaluation seemed to suffer 
from data leakage as the training and testing samples were allowed to be segmented from the same recording. 
A bowel motility detection model was proposed in \cite{ulusar2013real}. 
While the model allowed real-time abdominal sound analysis and result notification, it was only based on energy features which are very sensitive to noise. Since the neonatal bowel sounds are inherently weak, impact of noise is a major concern in this approach.
The model was later extended in \cite{ulusar2014recovery} by adding spectral features and introduction of \textit{Naive Bayesian} and minimum statistics. Although a decent overall accuracy was reported, the application of this model was still limited by the bowel sound detection sensitivity (35.23\% in total). 
The feasibility of using \textit{Higher Order Statistics} for bowel sound detection was discussed in \cite{sheu2014higher}. The method was proven effective and noise robust in synthetic data, but it performed with some limits in real-world collected data because it misclassified all non-peristalsis as peristalsis samples. 
Bowel sound recognition using Support Vector Machines (SVM) classifier was investigated in \cite{yin2018bowel}. The method provided a better-balanced sensitivity and specificity as compared to the aforementioned studies. Yet, the model was only evaluated on ideal laboratory data without any talking and equipment noises. The applied value of the model in real-world scenarios is yet to be explored.

Deep learning has shown great promise in many biomedical applications \cite{wainberg2018deep,sitaula2021attention}. Significant research is still underway to realize the full potential of deep learning in biomedicine and addressing its potential challenges \cite{han2020deep}. It was applied in the latest study on automated detection of bowel sounds \cite{liu2018bowel}. The method used a human voice recognition algorithm based on \textit{Mel Frequency Cepstrum Coefficient} (MFCC) features and \textit{Long-Short Term Memory} (LSTM) \cite{hochreiter1997long} neural network was applied to bowel sound detection. This novel attempt was theoretically justified and experimentally proven feasible. However, audio recordings were segmented into 0.1-second pieces for processing in this work. It seems less likely for LSTM to extract useful time dependency in such short segments.  

Despite the progresses made, none of the existing methods have proven 
feasible and effective for detection of neonatal bowel sound. There are no current benchmarks for objective neonatal peristalsis detection and characterisation. Due to the weakness of neonatal intestinal peristalsis, the interference of irregular breath and heart sound, and the various environment and equipment noises in NICU \cite{lahav2015questionable}, abdominal sounds recorded from newborns are usually of poor quality, which presents unique challenges to identify and characterize bowel sounds. 

In this work, we propose an effective method to automatically identify and locate bowel sounds in neonatal abdominal auscultatory recordings, which is a key step towards bowel condition diagnosis.
Specifically, our method leverages both Deep Learning (DL) and Laplace Hidden semi-Markov Model (HSMM) for the neonatal bowel sound classification. 
\textcolor{black}{Here, the DL model provides the probabilistic outputs based on the high-order semantic information present in the acoustic signals, whereas the HSMM uses these probabilistic outputs and temporal patterns to provide the optimized linear sequence labeling information. This complementary components altogether result in the improvement of classification performance significantly.
}
Evaluation on real neonatal bowel sound dataset shows that our method produces a superior result (89.81\%) compared to the second-best method (86.86\%) in terms of accuracy.
To the best of our knowledge, this is the first study on automated ``neonatal" bowel sound detection. 

\textcolor{black}{In summary, the main \textbf{contributions} of this paper are as follows:
\begin{itemize}
    \item[(1)] design of a novel CNN architecture to classify the neonatal bowel sounds into two classes (P and NP), which, we believe, is the first study;
    \item[(2)] optimize the probabilistic outputs of the proposed CNN model to improve the performance further using the Laplace HSMM model; and
    \item[(3)] demonstrate the superiority of the proposed approach (CNN model+Laplace HSMM) against several well-established popular ML methods and other DL methods using popular evaluation measures.
\end{itemize}
}

The rest of this paper is organized as follows. Section \ref{sec:method} describes our neonatal bowel sound detection method in detail. Section \ref{sec:result} presents the experimental results. Discussion is given in Section \ref{sec:discussion}. Section \ref{sec:conclusion} concludes the entire work.

\section{Methods} \label{sec:method}
This section presents our methods for data collection, signal processing, neonatal peristalsis sound detection, and model evaluation. Specifically, the peristalsis sound detection method consists of two modules: 
CNN for initial sound segment classification and a Laplace HSMM for classification optimization. 

\subsection{Data Acquisition and Annotation}
A total of 49 newborn infants admitted to our tertiary NICU for management of prematurity were recruited for the study during the period from February 2018 to September 2018. The newborn infants were selected based on the following inclusion criteria: 
1) preterm or term infants on full feeds $\textgreater$ 120~mL/kg/day; 2) infants not critically unwell, tolerating full enteral feeds and without any known diagnosis of bowel disease; 3) infants on continuous positive pressure ventilation (CPAP). 
Newborn infants who were intubated, ventilated, on mechanical ventilation, requiring any inotropes, or having significant chromosomal anomalies were excluded from this study. 
Informed consent was obtained from each study participant's parent. The study received human ethics approval from Western Sydney Local Health District Human Research Ethics Committee on 12 Apr 2018 (LNR/17/WMEAD/516).

\begin{figure}[tb]
	\centering
	\includegraphics[width=0.95\columnwidth]{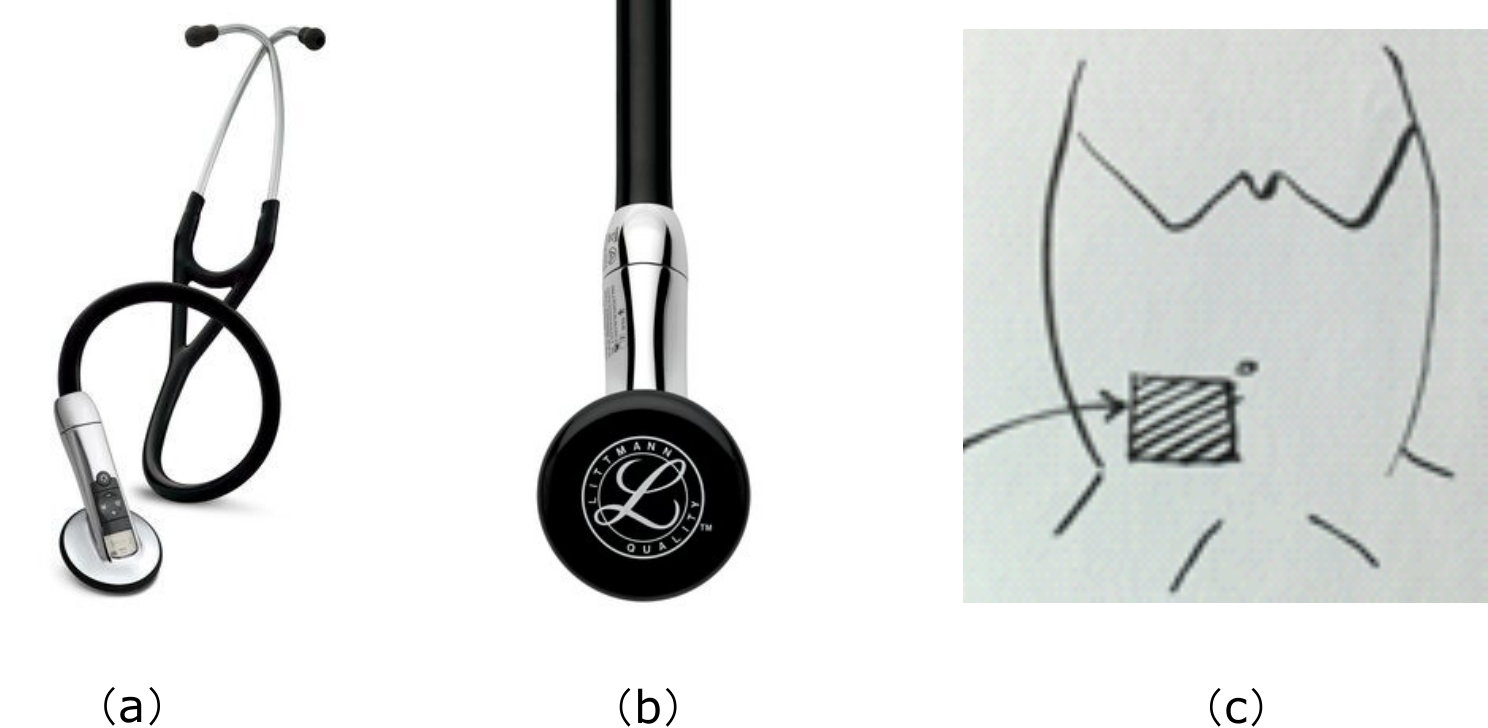}
	\caption{
		Abdominal sound auscultation device and position. 
		(a) 3M\textsuperscript{\texttrademark} Littmann\textsuperscript{\textregistered} Electronic Stethoscope Model 3200. 
		(b) Diaphragm of the digital stethoscope. 
		(c) The right lower quadrant abdominal wall.
	} 
	\label{fig:ds}
\end{figure}

The $3M^\text{TM} Littmann$ digital stethoscope (Model 3200 with Bluetooth, 3M, USA) \cite{3M20183M} was used for abdominal sound auscultation in this study. We used the diaphragm mode which ampliﬁes the sounds from 20 - 2000Hz, but emphasizes the sounds between 100 - 500Hz. For each recruited infant, who was placed supinely on the bed, an uninterrupted 60-second abdominal sound recording was captured by applying the diaphragm of the digital stethoscope on the abdominal wall in the right lower quadrant (RLQ). The rationale behind this is that the abdominal sound from the RLQ area is less interfered by the heart and lung sound. 
Fig. \ref{fig:ds} shows the used digital stethoscope and the auscultation position. 

The recorded abdominal sounds were transmitted to a research laptop via Bluetooth for visual interpretation and annotation. 
Annotation was created and validated on ELAN \cite{nijmegen2020elan}, a research tool specially designed for annotating audio and video data, by our neonatologists to indicate the onset and duration of bowel sound, heart sound, and various NICU noises in a recording. Simultaneous ultrasound was performed to guide annotation.

\subsection{Preprocessing}
Each annotated abdominal sound recording was sliced into overlapping segments by applying a 6-second length rectangular window function. In our ablative study in section \ref{segment_length}, segment length of 6s yielded the best results.  The step between the onsets of successive windows was set to 0.1 second. The 6-second segment length is long enough to provide useful insights for neonatologists to understand the bowel condition and also short enough for achieving a precise location of bowel sound in a recording.
After segmentation, a total of 16,401 abdominal sound segments were obtained. 
Those segments containing bowel sound were labeled as P (peristalsis), whereas the others were labeled as NP (non-peristalsis). The numbers of P and NP segments were 14,410 and 1,991, respectively. 

We calculated the MFCC features to represent each abdominal sound segment. MFCC is commonly used in automatic human speech recognition \cite{ittichaichareon2012speech}. It takes into account human perception sensitivity at appropriate frequencies by converting the conventional frequency ($f$) to Mel Scale ($M(f)$). The conversion rule is shown in Equation \ref{equ:mel}.
\begin{equation} \label{equ:mel}
M(f)=1125 \ln (1+f / 700)
\end{equation} 
The reason we use MFCC on abdominal sound is that the auscultation is highly dependent on medical expert perception. Representing abdominal sound with MFCCs facilitates development of an automated model to interpret the abdominal sounds in the similar way auscultated by medical experts.
More importantly, it has been proven that abdominal sounds are very similar to human speech in terms of the predictablility of the changes of spectrum versus time \cite{liu2018bowel}. 

The hyper-parameters setting in calculation of MFCC can be found in Table \ref{tab:parameters}. A mean operation is performed to summarize the calculated coefficients in each time-frame to obtain 
\textcolor{black}{a sequence of 24 MFCC values in one dimension}, 
which will be the input to the proposed bowel sound detection model. This segment length produced the best results in our ablative study of various sequence lengths described in section \ref{seq_length}.

\begin{table} [tb]
	\centering
	\caption{The hyper-parameters used in calculation of MFCCs.}
	\label{tab:parameters}
	\begin{tabular}{p{1cm}|p{5cm}|p{1cm}} 
		\hline 
		\hline 
		Parameter & Description & Value \\
		\hline  
		sfreq & the sampling rate of abdominal sound & 4000 \\
		\hline 
		winlen & the length of the analysis window & 0.025s \\
		\hline 
		winstep & the step between successive windows & 0.01s \\
		\hline 
		numcep & the number of cepstrum & 24 \\
		\hline 
		nfilt & the number of filters in the filter bank & 26 \\
		\hline 
		nfft & the Fast Fourier Transform size & 512 \\
		\hline 
	\end{tabular}

\end{table}

\begin{figure}[tb]
	\centering
	\includegraphics[height=135mm, width=0.45\textwidth,keepaspectratio]{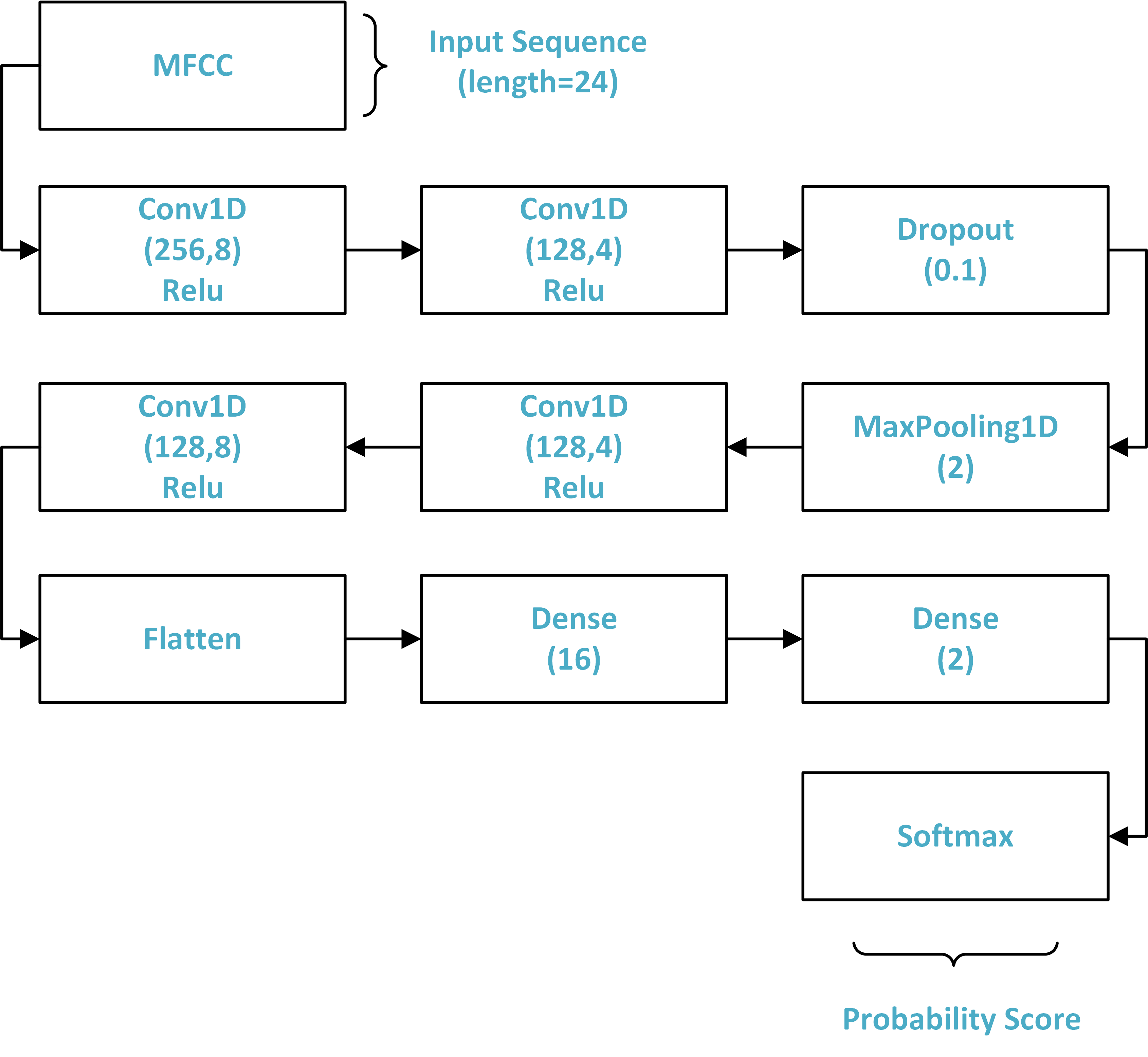}
	\caption{
		\textcolor{black}{Architecture of the proposed 
		convolutional neural network with 11 layers, where $Conv1D(x, y)$ denotes the 1D convolutional layer with number of filters as $x$ and the kernel size as $y$. Similarly, $Dropout(0.1)$, $MaxPooling1D(2)$, and $Dense(x)$ denote the dropout rate as $0.1$, max pooling size as $2$, and number of units in dense layer as $x$, respectively. }
 	} 
	\label{fig:mcnn}
\end{figure}

\subsection{Convolutional Neural Network-based Classification}
Convolution is useful in learning high-level representations of data, which is commonly applied in time series, image, and video data \cite{lecun1995convolutional}.
In this work, we propose a 
CNN 
to distinguish peristalsis from non-peristalsis sound.
The architecture of the proposed model is shown in Fig. \ref{fig:mcnn}.
Our proposed model accepts an MFCC sequence of length 24 as input. The most important part of CNN is the single channel convolution block, in which convolution
\textcolor{black}{of kernel size of 8 and number of filters of 256} at first 
applied to the input sequence to extract high-level features. We choose 8 as the best kernel size in our work from the empirical study (refer to Table \ref{tab:ablative_1dcnn}).  
The convolution block is specially designed to capture the burst characteristic of bowel sound. It has been proven that the bursting bowel sound can lead to an extremely uneven energy distribution in an abdominal sound recording along the time axis \cite{allwood2018advances}. Since the mel-frequency cepstrum (MFC) is a representation of the short-term power spectrum, the uneven energy distribution will also be reflected in MFCC accordingly. 
Hence, by applying 
the convolutions with an appropriate kernel capturing subtle energy changes in successive time frames, 
our proposed CNN model can learn useful energy patterns
to effectively characterize bowel sound.
A total of 
\textit{4 Convolutional} layers are included in our proposed model. Each convolution operation is followed by a \textit{ReLU} activation to enable non-linear transformation. 
\textit{Dropout} is used to overcome the problem of over-fitting
and \textit{Max Pooling} layer is applied to 
reduce the dimension and achieve the spatial invariant feature maps. Furthermore, inspired by the recent CNN architecture proposed by Anders et al. \cite{anders2020comparison}, which has suggested to use Convolutional layers after Max pooling layer, we adopt a similar approach to order in our work to capture the discriminating semantic information. 
The learned patterns achieved through several intermediate layers (e.g., Conv1D, Max pooling, etc.) impart hierarchical semantic information related to the input sequence, MFCC.
Eventually, the network estimates a probability distribution over peristalsis and non-peristalsis classes with the \textit{Softmax} function \cite{goodfellow2016deep}. The detailed hyper-parameters used in our work are presented in Table \ref{tab:hyper_parameters_1dcnn}. 
\textcolor{black}{To tune the hyper-parameters in this study, we first find the best optimizer in terms of higher classification accuracy keeping other parameters fixed (e.g., train/validation=0.3, lr=0.01, decay=1e-03, number of epochs=200 and batch size=32), which are chosen randomly. After the identification of the best optimizer with the highest classification accuracy, we proceed to select the best learning rate (lr) keeping other remaining parameters fixed. We continue this process repeatedly and, finally, find all optimal hyper-parameters in this study. }

\begin{figure}[tb]
	\centering
	\includegraphics[width=0.95\columnwidth,height=100mm,keepaspectratio]{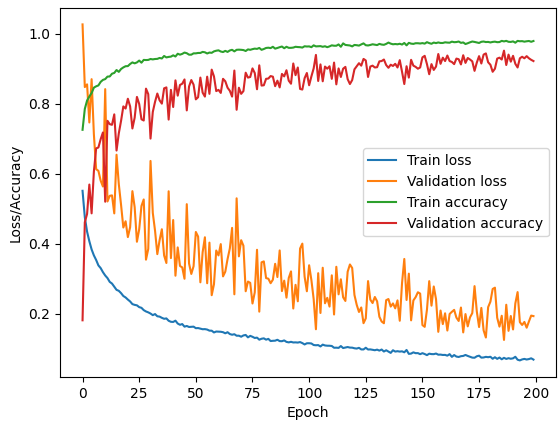}
	\caption{Sample train/validation accuracy and loss curve produced by our proposed CNN model for bowel sound classification
	} 
	\label{fig:test_test_evaluation}
\end{figure}

\textcolor{black}{
All of our experiments are performed in leave-one-patient-out CV (LOPOCV) approach, where one subject is considered as testing split and remaining as training split.
To train the DL models, we design a validation set by further splitting training set into training and validation splits with 70/30 ratio. With the help of the validation accuracy and losses, we tune the DL models. Based on such trained model, we, finally, evaluate the testing accuracy of the testing subject and repeat this process.
To train other machine learning models, we evaluate the testing accuracy by using the LOPOCV approach alone and repeat this process.
Both approaches (DL and non-DL) produce the testing accuracy for the comparison purpose.}
Our proposed model produces the best-fit convergence during bowel sound classification (refer to Fig. \ref{fig:test_test_evaluation}).


\begin{table}[]
    \centering
    \caption{Hyper-parameter settings used for training the deep learning (DL) models in our work.}
    \begin{tabular}{p{5cm}|p{2cm}}
    \hline
    \hline
     Hyper-parameters & Value\\
     \hline
         Optimizer \textcolor{black}{('Adam','RMSProp','SGD')}& RMSProp \\
         \hline
         Learning rate \textcolor{black}{(1e-01,1e-05, 1e-06)}& 1e-05 \\
         \hline
         Decay \textcolor{black}{(1e-01, 1e-06, 1e-08)}& 1e-06 \\
         \hline
         Epochs \textcolor{black}{(100, 200, 300)}& 200 \\
         \hline
         Batch size \textcolor{black}{(8, 16, 32)}&16 \\
         \hline
         Train/validation split \textcolor{black}{(0.2, 0.3, 0.4)} &0.3 (70/30) \\
         \hline
    \end{tabular}
    \label{tab:hyper_parameters_1dcnn}
\end{table}

\subsection{Laplace Hidden Semi-Markov Model Refinement}

Abdominal sound changes over time, but the change is not completely random because peristalsis is a naturally durative process. For example, it is less likely to find a sudden pause within a continuous peristalsis or to find a transient peristalsis which just lasts for 0.1s. 
Therefore, for an abdominal sound recording, if we arrange all segments in time order, the transition between the segment states, i.e. peristalsis or non-peristalsis, should be subject to certain probability distribution.
Motivated by this fact, a HSMM is applied to model the segment transition process to refine the bowel sound classification result. 
HSMM is an extension to Hidden Markov Model (HMM) \cite{rabiner1989tutorial}. They both have been proven to be useful in modeling repetitive events by considering the timing and the temporal order of events \cite{yu2010hidden}. As compared to HMM, HSMM allows a hidden state to have a duration distribution, which is more suitable to be applied on abdominal sound with durative peristalsis.

\begin{figure}[tb]
	\centering
	\includegraphics[width=0.90\columnwidth]{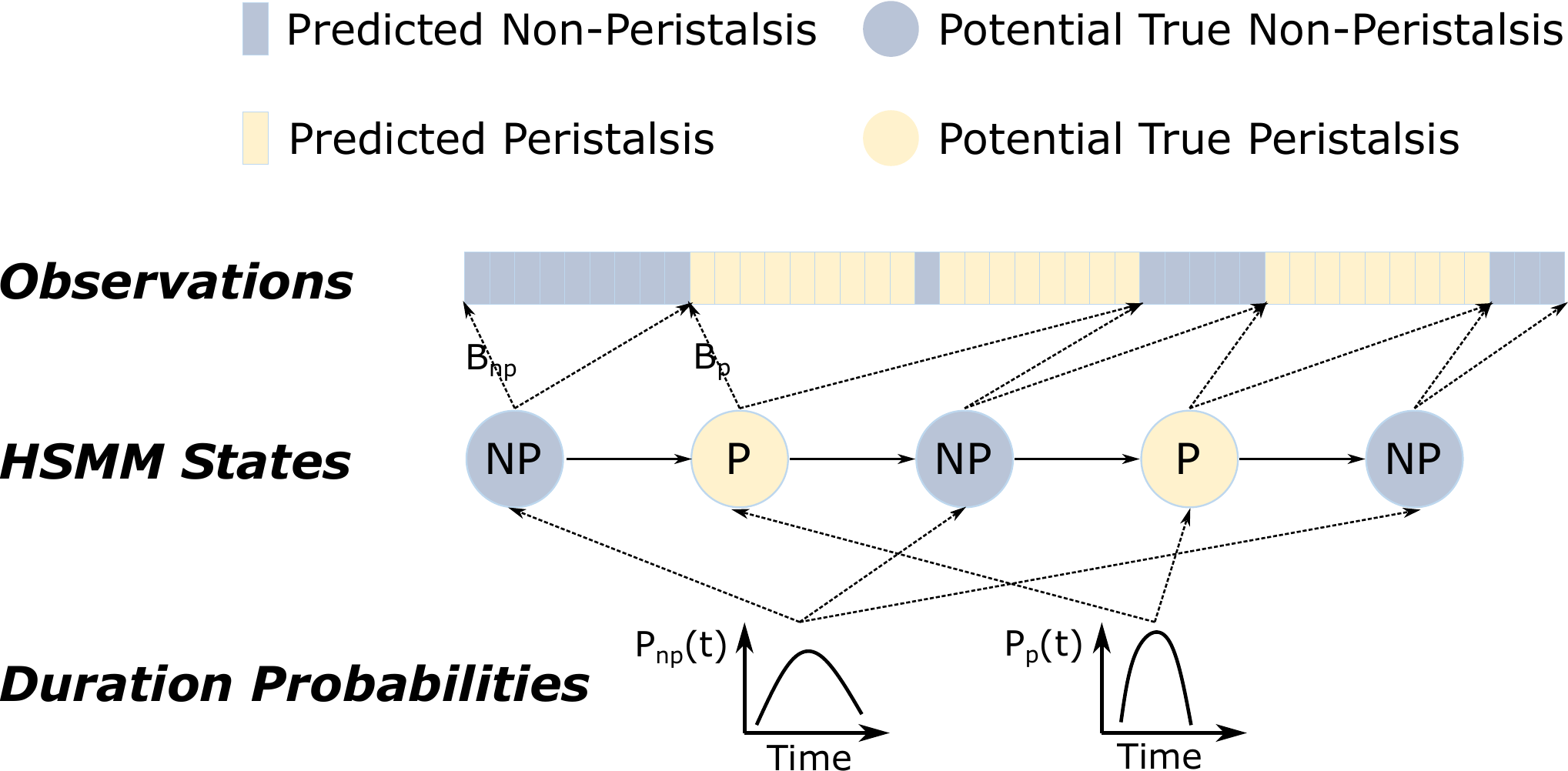}
	\caption{
		Overview of the HSMM modeling process. 
		Peristalsis and non-peristalsis segments are represented by the yellow and grey blocks, respectively. 
		$B_{np}$ and $B_p$ denotes the emission probability of the hidden NP and P states, respectively.
		$P_{np}$ and $P_p$ denotes the duration probability of the hidden NP and P states, respectively.
	} 
	\label{fig:hsmm}
\end{figure}

The HSMM modeling process is presented in Fig. \ref{fig:hsmm}. Segments from an abdominal sound recording arranged in time order serve as the observation sequence. Labels of the segments are predictions from the CNN model. 
The HSMM states denote the potential true labels. Duration of a state represents the number of observations produced while in the state.
The refinement is achieved by estimation of the HSMM state sequence. 

Formally, let $O_{1: T} = O_{1}, \ldots, O_{T}$ denote the observation sequence, where $O_{n} \in [0, 1]$ is the predicted probability of the $n$-th segment belonging to class P. 
Let $S_{1: N} = (j_{1}, d_{1}), \ldots, (j_{N}, d_{N})$ denote the HSMM state sequence, in which \textcolor{black}{$j_{n} \in \{P, NP\}$} and $d_{n}$ 
\textcolor{black}{represent the state label and duration of the $n$-th hidden state}, respectively.
The objective function is defined as
\begin{equation}
\operatorname{argmax}_{S_{1: N}} P(S_{1: N} \mid O_{1: T}, \lambda),
\end{equation}
where $\lambda$ denotes the model parameters including initial probability, transition probability, emission probability, and duration probability. 

The initial state distribution is determined based on the percentages of class P and NP being the first state in the training recordings. It is given by
\begin{equation}
\pi_{j} = 
\begin{cases}
	\frac{Supp(P_{ini})}{n}, & j = P\\
	1 - \pi_{P}, & j = NP	
\end{cases}, 
\end{equation}
where $Supp(P_{ini})$ denotes the number of training recordings with class P as the initial state and $n$ represents the amount training recordings. 

The types of the HSMM states are assumed to change in every transition. 
So, the state transition probability is given by 
\begin{equation}
a_{i j}=1, \quad  i \neq j \ \& \ i, \textcolor{black}{j \in \{P, NP\}}.
\end{equation}

{We use the \textit{Laplace distribution} to provide a continuous modeling of the state emission probabilities. 
As a result, we are able to capture them for both P and NP events that are continuous in bowel sound.}
\textcolor{black}{Here, we have observed experimentally that emission probabilities could be properly modelled by Laplace distribution, based on this dataset.} 
For a given state \textcolor{black}{$j \in \{P, NP\}$}, the emission probability distribution is: 
\begin{equation}
b_{j}\left(x|\mu, \sigma\right)=\frac{1}{2 \sigma} \exp \left(-\frac{|x-\mu|}{\sigma}\right),
\end{equation}
where $\sigma$ denote the standard deviation, \textcolor{black}{$x$ the output of the corresponding class node in the softmax output layer of the CNN} and $\mu$ is the mean value. We set $\mu = 1$ if $j = P$, otherwise $\mu = 0$.
As compared to the conventional discrete emission probabilities, which are learned from training confusion matrix, the Laplace distribution 
\textcolor{black}{provide the} results improvements for the network output ($x$). Experimental evidences can be found in Section \ref{subsec:lhsmm}. To be distinguished from the conventional practice, we refer to the modified HSMM as Laplace HSMM.

The duration probability of each state is learned from the training data. We use $p_{j}(d)$ to represent the probability of state $j$ having duration $d$, where $j \in\{P, NP\}$. Algorithm \ref{alg:getduration} presents the learning process, in which the duration frequency counting is performed in lines 10-17. \textcolor{black}{Laplace} smoothing is performed in lines 23-27 to avoid discontinuous duration probabilities.  

\begin{algorithm} 
	\label{alg:getduration}
	\caption{Duration Probability Matrix Learning}
	\LinesNumbered
	\KwIn{
		$trainSubjects$\;
	}
	\KwOut{$p_{j}(d)$\;}
	
	$durMtx$ $\gets$ new(dictionary) \;
	\For{each $recording$ in $trainSubjects$}{
		$segments$ $\gets$ segmentation($recording$)\;
		$segments$ $\gets$ sort($segments$)\;	
		current label $clb \gets P$\;
		$d \gets 0$\;
		\For{$seg$ in $segments$}{
			\If{$seg.label$ = $clb$}{
				$d \gets d + 1$\;
			}
			\Else{
				\If{d != 0}{
					\If{durMtx[clb, d] exist}{
						$durMtx[clb, d] \gets durMtx[clb, d] + 1$\;
					}
					\Else{
						$durMtx[clb,d] \gets 1$\;
					}
					$clb$ $\gets$ $seg.label$\;
					$d \gets 1 $
				}
			}
		}
		\If{d != 0}{
			\If{durMtx[clb, d] exist}{
				$durMtx[clb, d] \gets durMtx[clb, d] + 1$\;
			}
			\Else{
				$durMtx[clb, d] \gets 1$\;
			}
		}
	}
	\For{$j$ in \textcolor{black}{$\{P, NP\}$} and $d$ in [0, len($durMtx[j]$)]}{
		\If{durMtx[j, d] exist}{
			$durMtx[j, d] \gets durMtx[j, d] + 1$\;
		}
		\Else{
			$durMtx[j, d] \gets 1$\;
		}
	}
	\For{$j$ in \textcolor{black}{$\{P, NP\}$} and $d$ in len($durMtx[j]$)}{
		$p_{j}(d) \gets durMtx[j, d] / \sum(durMtx[j])$\;
	}
	\Return $p_{j}(d)$\;
\end{algorithm}

We use the Viterbi algorithm \cite{forney2005viterbi} for the maximum likelihood estimation of the hidden state sequence $S_{1: N}$. The algorithm is popularly used in HMM. We modify it as follows to be compatible with the Laplace HSMM. Firstly, we define $\delta_{j}(t)$ as the highest probability of a hidden state sequence of length $n$, which accounts for the observations of length $t$ and ends in state $j$:
\begin{equation}
\delta_{j}(t)=\max _{ S_{1:n-1}} P \left( S_{1:n} \mid O_{1:t}, \lambda\right), S_{n} = (j, d_{n}).
\end{equation}

Then we have
\begin{equation}
\delta_{j}(1)=\pi_{j}p_{j}(1)b_{j}\left(O_{1}\right),
\end{equation}
and 
\begin{equation}
\delta_{j}(t)=\max _{d} \max _{i} \left\{ \delta_{i}(t-d) a_{i j}p_{j}(d) \prod_{\tau=t-d+1}^{t} b_{j}\left(o_{\tau}\right)\right\},
\end{equation}
\textcolor{black}{where $o_{\tau}$ is the outcome of the Laplace distribution function that took the output of the CNN at timestep $\tau$ as its input.}
Finally, the most likely HSMM state sequence can be obtained from backtracking the following equation 
\begin{equation}
\max_{S_{1: N}} P(S_{1: N} \mid O_{1: T}, \lambda)=\max _{j} \delta_{j}(T).
\end{equation}

\begin{table*} [!t]
	\centering
	\caption{The Hand-engineered Features for Bowel Sound Detection}
	\label{tab:features}
	\begin{tabular}{p{2cm}|p{4cm}|p{10cm}} 
		\hline 
		\hline 
		Number & Feature & Description \\
		\hline  
		1 - 5 & Jitters & Measure of variations from true or target periodicity of a presumably periodic signal. Five kinds of jitters are extracted: local jitter, local and absolute jitter, rap jitter, ppq5 jitter, and ddp jitter. The definitions can be found in \cite{ppgb2011jitter}. \\
		\hline 
		6 - 11 & Shimmers & Measure of variations in signal amplitude between consecutive vibratory cycles. Six kinds of shimmers are extracted: local shimmer, local db shimmer, apq3 shimmer, apq5 shimmer, apq11 shimmer, and ddp shimmer. The definitions can be found in \cite{ppgb2003shimmer}.\\
		\hline 
		12 & Skewness & Measure of asymmetry of probability distribution. \\
		\hline 
		13 & Kurtosis & Measure of 'tailedness' of probability distribution function. \\
		\hline 
		14 - 19 & Wavelet Decomposition Subband Energy & Energy of each frequency subband decomposed by the wavelet transform. The decomposition level is set to 5. For signals sampled at 4000hz, the corresponding frequency subbands are: [0 - 62Hz], [62 - 125Hz], [125 - 250Hz], [250 - 500Hz], [500 - 1000Hz], and [1000 - 2000Hz].  \\
		\hline 
		20 - 45 & Spectral Subband Centroids & Frequency centroid of each frequency subband decomposed by the mel filterbank. The number of filters in the filterbank is 26 and the FFT size is 512. For signals sampled at 4000hz, the corresponding frequency subbands are: [0 - 74Hz], [36 - 113Hz], [74 - 155Hz], [113 - 199Hz], [155 - 245Hz], [199 - 293Hz], [245 - 344Hz], [293 - 398Hz], [344 - 454Hz], [398 - 513Hz], [454 - 575Hz], [513 - 641Hz], [575 - 710Hz], [641 - 782Hz], [710 - 858Hz], [782 - 938Hz], [858 - 1022Hz], [938 - 1110Hz], [1022 - 1203Hz], [1110 - 1300Hz], [1203 - 1403Hz], [1300 - 1511Hz], [1403 - 1624Hz], [1511 - 1743Hz], [1624 - 1868Hz], [1743 - 2000Hz] \\
		\hline 
		46 - 71 & Spectral Subband Energy & Energy of each frequency subband decomposed by the mel filterbank. The number of filters in the filterbank is 26 and the FFT size is 512. For signals sampled at 4000hz, the filtered frequency subbands are the same as above. \\
		\hline 
		72 - 97 & Log Subband Energy & Log energy of each frequency subband decomposed by the mel filterbank. The number of filters in the filterbank is 26 and the FFT size is 512. For signals sampled at 4000hz, the filtered frequency subbands are the same as above.  \\
		\hline 
		98 - 121 & MFCCs & Mel-frequency cepstral coefficients. The number of cepstrum to return is 24. \\
		\hline 
	\end{tabular}
\end{table*}

\begin{table} [!t]
	\centering
	\caption{Parameters of The Compared Machine Learning Models}
	\label{tab:model parameters}
	\begin{tabular}{p{2cm}|p{5cm}} 
		\hline 
		\hline 
		Classifier & Parameters \\
		\hline  
		KNN & Number of neighbors K = 3.  \\
		\hline 
		Linear SVM & Cost C = 0.25  \\
		\hline 
		RBF SVM & Kernel Coefficient Gamma = [0.1, 0.01, 0.001, 0.0001, 0.02]; Cost C = [0.04, 0.08, 0.10, 0.16, 1]  \\
		\hline 
		DT & Max tree depth: 5.  \\
		\hline 
		RF & Max tree depth: 5; Number of estimators: 50.  \\
		\hline 
		MLP & Regularization term: 0.1; Hidden layer size: 100; Activation: ReLU; Max iteration number: 500.  \\
		\hline 
		AdaBoost & Number of estimators: 50   \\
		\hline 
		NB & Not applicable.  \\
		\hline 
		QDA & Not applicable.  \\
		\hline 
	\end{tabular}
\end{table}

\subsection{Performance Evaluation}
We utilize leave-one-patient-out cross validation (LOPOCV) to evaluate the proposed model. For each round of LOPOCV, abdominal sound segments from one patient are used as test data while segments from the remaining 48 patients are used for training. 

\subsubsection{Baseline Methods}
We compare our method with multiple machine learning algorithms using hand-engineered signal features derived from the related literature, including jitters and shimmers \cite{kim2011estimation,kim2011non,kim2012awareness,yin2015bowel},  higher order statistics \cite{lin2013enhancing, sheu2014higher}, wavelet subband energies \cite{dimoulas2007long}, spectral centroids \cite{ulusar2014recovery}, spectral subband energies \cite{ranta2001wavelet,ulusar2014recovery}, and MFCCs \cite{liu2018bowel}. 
A total of 121 features were extracted, as summarized in Table \ref{tab:features}. The parameters of the feature extraction methods were fine-tuned to adjust to the experimental abdominal sound recordings and the extracted features were normalized to zero mean and unit variance. 
To reduce the risk of over-fitting, feature selection was applied to select the best 20 features for training and evaluation.  
Specifically, we followed the feature selection method proposed in \cite{will2020a}, which implements the following processing steps:
\begin{itemize}
	\item Remove features with more than 60\% missing values.
	\item Remove features with a single unique value.
	\item Remove collinear features with a \textcolor{black}{correlation} of 0.98.
	\item Rank features based on their importance scores calculated by a Gradient Boosting Decision Tree (GBDT). 
\end{itemize} 

The compared machine learning algorithms are K-Nearest Neighbors (KNN), Linear SVM, Radial Basis Function (RBF) SVM, Decision Tree (DT), Random Forest (RF), MLP, AdaBoost, Naive Bayes (NB), and Quadratic discriminant analysis  (QDA). The model parameters were determined experientially, which are summarized in Table \ref{tab:model parameters}.

\subsubsection{Evaluation Metrics}
We used classification accuracy (ACC), area under the \textcolor{black}{receiver operating characteristics (ROC)} curve (AUC) \cite{fawcett2006introduction}, macro-averaged F1 (MA\_F1), and weighted F1 (WT\_F1) as the evaluation metrics of overall model performance. 
Specifically, MA\_F1 is the unweighted average of F1 score of each class, whereas WT\_F1 is F1 average weighted by the number of instances of each class. 
Besides, recall (REC), precision (PRE), and F1 scores (F1) for each class are also measured to allow inspection of the model performance on individual classes. The metrics used in this study are formally defined with the following equations: 

\begin{equation}
ACC = \frac{TP+TN}{\sum},
\end{equation}
\begin{equation}
AUC = \int_{x=0}^{1} \mathrm{ROC}(x) d x,
\end{equation}
\begin{equation}
MA\_F1 = \frac{F1_{P}+F1_{NP}}{2},
\end{equation}
\begin{equation}
WT\_F1 = \frac{F1_{P} \times Supp(P) + F1_{NP}\times Supp(NP)}{\sum},
\end{equation}
\begin{equation}
PRE = \frac{TP}{TP+FP},
\end{equation}
\begin{equation}
REC = \frac{TP}{TP+FN},
\end{equation}
\begin{equation}
F1 = 2 \times \frac{PRE \times REC}{PRE + REC},
\end{equation}
where $TP$, $TN$, $FP$, and $FN$ denote true positive, true negative, false positive and false negative, respectively, and $\sum$ represents the amount of instances in the data set. 
\textcolor{black}{ROC(.) denotes the curve plot of true positive rate (y-axis) versus false positive rate (x-axis). }
$F1_{P}$ and $F1_{NP}$ denote the F1 scores of class P and NP, respectively, while $Supp(P)$ and $Supp(NP)$ denote the number of instances of class P and NP, respectively.

\begin{table*} [tb]
	\centering
	\scriptsize
	\caption{Bowel Sound Detection Results with Leave-one-patient-out Cross Validation}
	\label{tab:results}
	\begin{threeparttable}
		\begin{tabular}{p{3.0cm}p{0.9cm}p{0.9cm}p{0.9cm}p{0.9cm}p{0.1cm}p{0.8cm}p{0.8cm}p{0.6cm}p{0.1cm}p{0.8cm}p{0.8cm}p{0.6cm}}
			\hline
			\hline
			\specialrule{0em}{1pt}{2pt}
			\multirow{2}{*}{Method} & \multirow{2}{*}{ACC} & \multirow{2}{*}{AUC} & \multirow{2}{*}{MA\_F1} & \multirow{2}{*}{WT\_F1} && \multicolumn{3}{l}{Non-peristalsis} & & \multicolumn{3}{l}{Peristalsis} \\
			\specialrule{0em}{1pt}{1pt}
			\cline{7-9}
			\cline{11-13}
			\specialrule{0em}{1pt}{2pt}
			&&&&&& REC & PRE & F1 && REC & PRE & F1 \\
			\specialrule{0em}{1pt}{1pt}
			\hline
			\specialrule{0em}{1pt}{2pt}
             \textbf{Proposed} &  \textbf{0.8981} &  \textbf{0.8396} &  \textbf{0.7928} &  \textbf{0.9046} &&  0.7624 &  \textbf{0.5589} &  \textbf{0.6450} &&  0.9169 &  0.9654 &  \textbf{0.9405} \\
			
			\specialrule{0em}{1pt}{1pt}
			\hline
			\specialrule{0em}{1pt}{1pt}
			KNN & 0.8075 & 0.6173 & 0.6020 & 0.8186 && 0.3661 & 0.2778 & 0.3159 && 0.8685 & 0.9084 & 0.8880 \\
			\specialrule{0em}{1pt}{1pt}
			
			Linear SVM & 0.7941 & 0.7558 & 0.6636 & 0.8222 && 0.7052 & 0.3348 & 0.4540 && 0.8064 & 0.9519 & 0.8731 \\
			\specialrule{0em}{1pt}{1pt}
			
			RBF SVM 1 (0.02, 1)\tnote{1}  & 0.8668 & 0.6885 & 0.6883 & 0.8669 && 0.4530 & 0.4517 & 0.4524 && \textbf{0.9240} & 0.9244 & 0.9242 \\
			\specialrule{0em}{1pt}{1pt}

			RBF SVM 2 (0.0001, 0.16)  & 0.5391 & 0.6869 & 0.4847 & 0.6115 && \textbf{0.8820} & 0.1934 & 0.3172 && 0.4917 & \bf 0.9679 & 0.6522 \\
			\specialrule{0em}{1pt}{1pt}
			
			RBF SVM 3 (0.001, 0.1) & 0.7761 & 0.7451 & 0.6467 & 0.8086 && 0.7042 & 0.3125 & 0.4329 && 0.7860 & 0.9506 & 0.8605  \\
			\specialrule{0em}{1pt}{1pt}
			
			RBF SVM 4 (0.01, 0.08) & 0.8190 &  0.7905 & 0.6960 & 0.8424 && 0.7529 & 0.3771 & 0.5025 && 0.8282 & 0.9604 & 0.8894 \\
			\specialrule{0em}{1pt}{1pt}
			
			RBF SVM 5 (0.1, 0.04) & 0.8543 & 0.7011 & 0.6849 & 0.8598 && 0.4987 & 0.4164 & 0.4538 && 0.9034 & 0.9288 & 0.9159  \\
			\specialrule{0em}{1pt}{1pt}
			
			DT & 0.7739 & 0.6614 & 0.6090 & 0.8013 && 0.5128 & 0.2716 & 0.3551 && 0.8100 & 0.9233 & 0.8629 \\
			\specialrule{0em}{1pt}{1pt}
			
			RF & 0.8556 & 0.7408 & 0.7067 & 0.8649 && 0.5892 & 0.4308 & 0.4977 && 0.8924 & 0.9402 & 0.9157 \\
			\specialrule{0em}{1pt}{1pt}
			
			MLP & 0.8520 & 0.6706 & 0.6648 & 0.8545 && 0.4309 & 0.3987 & 0.4142 && 0.9102 & 0.9205 & 0.9153  \\
			\specialrule{0em}{1pt}{1pt}
			
			AdaBoost & 0.8371 & 0.7320 & 0.6866 & 0.8511 && 0.5932 & 0.3882 & 0.4693 && 0.8709 & 0.9394 & 0.9038 \\
			\specialrule{0em}{1pt}{1pt}
			
			NB & 0.7014 & 0.6991 & 0.5833 & 0.7513 && 0.6961 & 0.2441 & 0.3615 && 0.7022 & 0.9436 & 0.8052 \\
			\specialrule{0em}{1pt}{1pt}
			
			QDA & 0.8143 & 0.6130 & 0.6024 & 0.8222 && 0.3471 & 0.2837 & 0.3122 && 0.8789 & 0.9069 & 0.8927 \\
			\specialrule{0em}{1pt}{1pt}
			\hline
		\end{tabular}
		\begin{tablenotes}
			\footnotesize
			\item[1] (x, y) denotes that the Gamma and Cost parameters of the SVM are set to x and y, respectively.
		\end{tablenotes}
	\end{threeparttable}
\end{table*}

\section{Experimental Results} \label{sec:result}
In this section, we present experimental results to demonstrate the effectiveness of our bowel sound detection model. We also show that the proposed Laplace HSMM strategy can introduce general improvements to achieve a more accurate bowel sound detection. 

\subsection{Comparative Results}
The comparative results are presented in Table \ref{tab:results}. It can be observed that the proposed CNN + HSMM method achieved the best scores on 7 out of 10 metrics in total, with the remaining 3 metrics being comparable to the best performing baselines. This suggests that the proposed method is effective in neonatal bowel sound detection and has introduced significant improvements to current methods.

The RBF SVM 1 obtained the second best overall classification accuracy and the highest recall rate of peristalsis samples. However, this is at the cost of misclassification of more than half the non-peristalsis samples with the recall value being just 0.4530. By contrast, the corresponding recall of the proposed method is 0.7624, which is significantly better. In practice, accurately recognizing the segments without bowel sound is very important because it helps to avoid unnecessary tests and treatments, and reduce risks for patients. The best recall of non-peristalsis samples was achieved by RBF SVM 2, but meanwhile the model obtained the lowest ACC, AUC, MA\_F1, WT\_F1, and peristalsis recall. This means that the model could hardly detect peristalsis samples.  

Given that the experimental dataset is very imbalanced, with 88\% of samples being peristalsis, the overall accuracy can not provide enough insights of the true model performance. A decent accuracy value can be easily obtained by classifying all samples as the dominated class. Therefore, the AUC score is more critical in measuring the model overall performance because it is very sensitive to class imbalance. It is worth to note that our model is the only one which achieves an AUC score close to 0.84, outperforming the second best by nearly 5\%. 
Besides, the MA\_F1, WT\_F1, and the individual class metrics also suggest that the proposed method outperformed the baseline algorithms.

\subsection{Ablative Study of Different Kernels}
In this subsection, we studied the efficacy of different kernels (2, 4, 8, and 16) based on fixed-sized input sequence of 24 and segment length of 6 s in our proposed CNN model. Detailed experimental results of classification performance of each kernel are presented in Table \ref{tab:ablative_1dcnn}. While observing the table, we notice that the best kernel size is 8 in our work. This work reveals that lower kernel size is unable to capture the semantics of signal information accurately, whereas the higher kernel size could repeat the discriminating information, which, in result, degrades the classification performance.

\begin{table}[tb]
    \centering
    \scriptsize
         \caption{Ablative study of four different channels (2, 4, 8, and 16) used in our work. Note that we use Laplace HSMM on top of the result produced from each deep learning model.}
    \begin{tabular}{c|c|c|c|c}
    \hline
    \hline
         Metrics & 2 & 4 & 8 & 16\\
         \hline
         Acc.&0.8473 &0.8846 &\bf 0.8981 &0.8801\\
         AUC&0.7441 &0.8010&\bf 0.8396&0.8071 \\
         MA\_F1&0.7009 &0.7626&\bf 0.7928&0.7600 \\
         WT\_F1&0.8594 &0.8915&\bf 0.9046&0.8886 \\
         NP\_REC&0.6077 &0.6906&\bf 0.7624&0.7107 \\
         NP\_PRE&0.4127 &0.5187&\bf 0.5589&0.5046 \\
         NP\_F1&0.4916 &0.5924&\bf 0.6450&0.5902 \\
         P\_REC&0.8805 &0.9115&\bf 0.9169&0.9036 \\
         P\_PRE&0.9420 &0.9552&\bf 0.9654&0.9576 \\
         P\_F1&0.9102 &0.9328&\bf 0.9405&0.9298 \\
         \hline
    \end{tabular}
    \label{tab:ablative_1dcnn}
\end{table}



\subsection{Comparison with Other DL Methods}
\begin{figure}[tb]
    \centering
    \includegraphics[height=50mm, width=\textwidth,keepaspectratio]{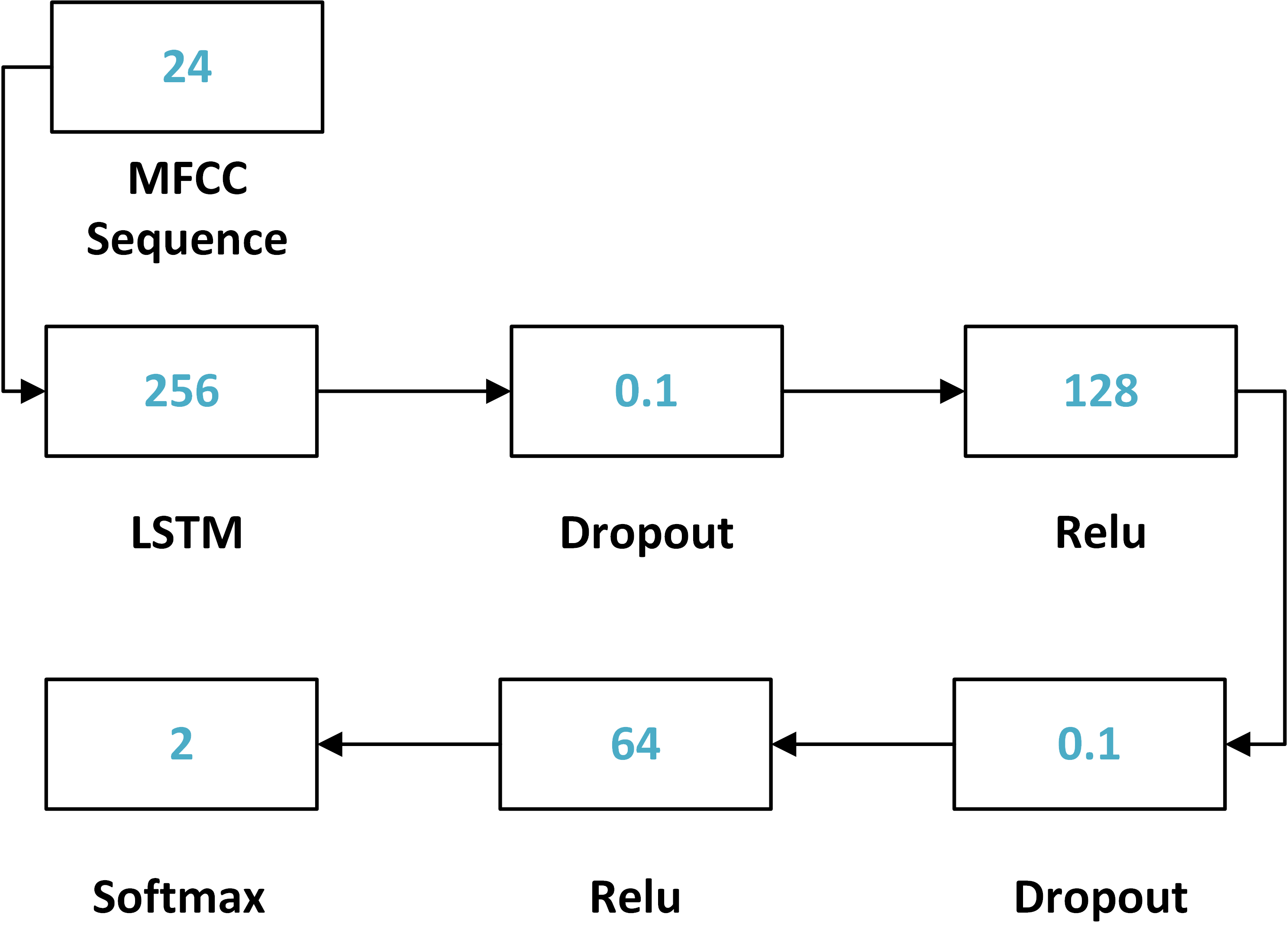}
    \caption{LSTM model used to compare our proposed CNN model.}
    \label{fig:lstm}
\end{figure}

\begin{figure}[tb]
    \centering
    \includegraphics[height=50mm, width=\textwidth,keepaspectratio]{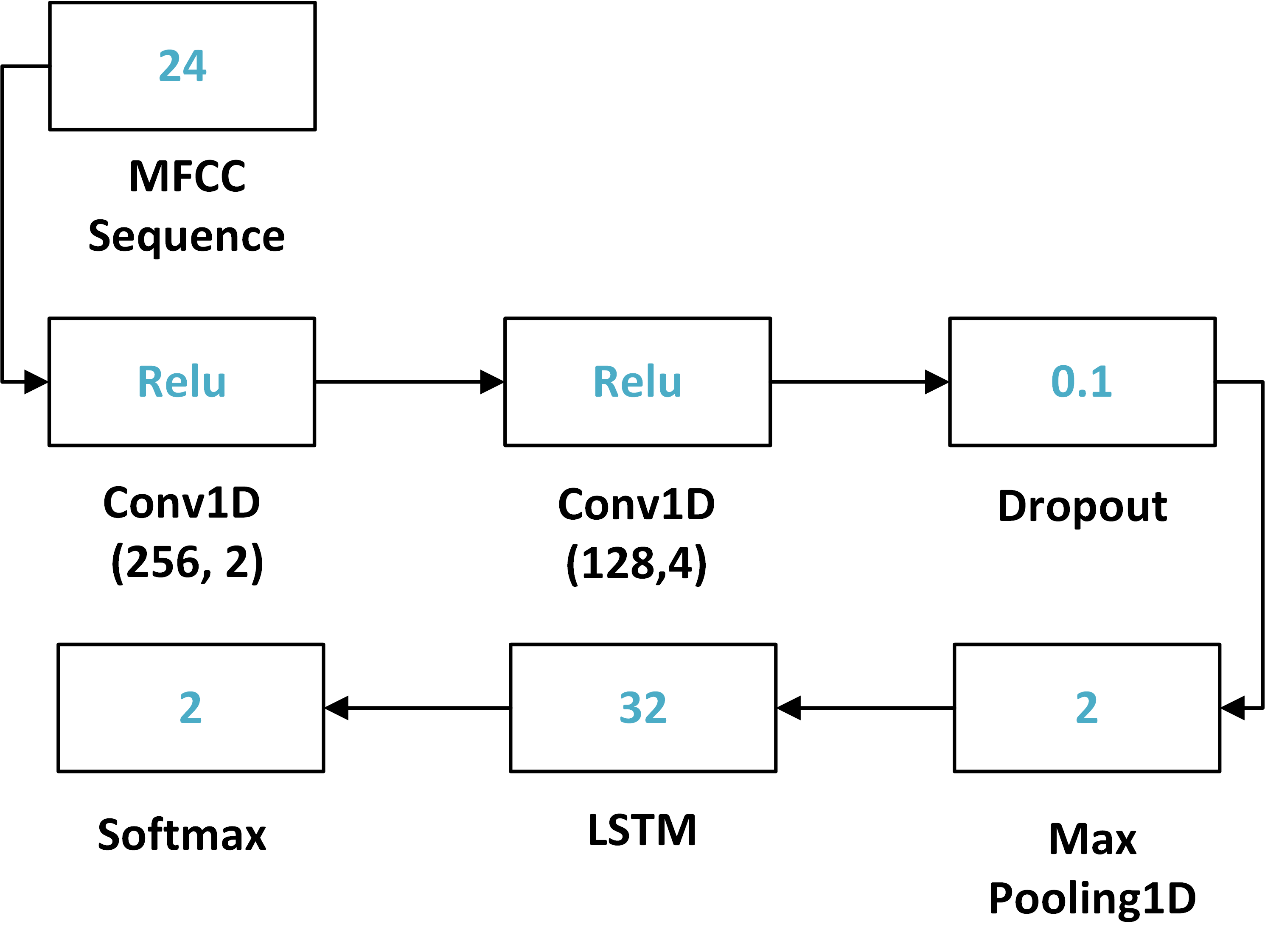}
    \caption{CNN-LSTM model used to compare with our proposed CNN model}
    \label{fig:cnn_lstm}
\end{figure}

Since there are no well-established benchmark datasets and DL models to classify neonatal bowel sound features, we compared our method against two different variants of DL methods: LSTM (Fig. \ref{fig:lstm}) and CNN-LSTM (Fig. \ref{fig:cnn_lstm}) using fixed-sized input sequence of 24 and segment length of 6 s. Experimental results, which are obtained after Laplace HSMM, are presented in Table \ref{tab:dl_models_comparison}. While observing the table, we notice that our proposed CNN model outperforms other two versions of DL models. 

\begin{table}[tb]
    \centering
    \scriptsize
         \caption{Comparison of our method against two DL models}
    \begin{tabular}{c|c|c|c}
    \hline
    \hline
         Metrics & CNN  &LSTM &CNN-LSTM\\
         \hline
         Acc.& \bf 0.8981 &0.6644 &0.7888 \\
         AUC& \bf 0.8396  &0.6966 &0.7794 \\
         MA\_F1&\bf 0.7928  &0.5612 &0.6684 \\
         WT\_F1& \bf 0.9046  &0.7223 &0.8197 \\
         NP\_REC&\bf 0.7624  &0.7393 &0.7670 \\
         NP\_PRE&\bf 0.5589  &0.2280 &0.3374 \\
         NP\_F1&\bf 0.6450  &0.3485 &0.4686 \\
         P\_REC&\bf 0.9169  &0.6541 &0.7919 \\
         P\_PRE& \bf 0.9654  &0.9478 &0.9609 \\
         P\_F1&\bf 0.9405 &0.7740 &0.8683\\
         \hline
    \end{tabular}
    \label{tab:dl_models_comparison}
\end{table}

\textcolor{black}{Furthermore, we compared our method against two baseline methods (LSTM and CNN-LSTM) using statistical analysis. First, we performed the normality test using Jarque-Bera (JB) test, which showed that our data were not normally distributed. Therefore, we used nonparametric Wilcoxon signed rank test to assess the statistical significance \cite{demvsar2006statistical} of our results compared to two baseline methods separately.
While comparing the performance of our method against LSTM and CNN-LSTM at subject-level (for 49 subjects), we found that our method achieves significantly higher weighted PRE, weighted REC, WT\_F1 and ACC, with the p-value of 1.181e-3, 4.727e-3, 1.56e-3, and 4.727e-3 respectively, compared to LSTM and with p-values of 1.9e-4, 1.516e-05, 7.739e-06, respectively, compared to  CNN-LSTM. 
Therefore, our proposed method produces an encouraging result against the baselines.
}

\subsection{Ablative Study of Varying Sequence Lengths}\label{seq_length}
In this subsection, we studied the relationship between sequence length and classification performance in our work. For this, we utilized 5 different sequence lengths (8, 16, 20, 24, and 26) with the fixed segment length of 6 s and evaluated our model. The experimental results, which are obtained after Laplace HSMM, are presented in Table \ref{tab:ablative_var_sequence}. The experimental results show that the best sequence length is 24. Therefore, we stipulate that the sequence length of 24 is sufficient to capture the discriminating information for bowel sound classification. 

\begin{table}[]
    \centering
    \scriptsize
         \caption{Comparative study of 5 different sequence length with our method. }
    \begin{tabular}{c|c|c|c|c|c}
    \hline
    \hline
         Metrics&8&16&20&24&26\\
         \hline
         Acc.&0.7729  &0.8156 &0.8817 &\bf 0.8981 &0.8878 \\
         AUC &0.7417 &0.7457 &0.8325 &\bf 0.8396 &0.8294  \\
         MA\_F1 &0.6433 &0.6756 &0.7710 &\bf 0.7928 &0.7769 \\
         WT\_F1&0.8061 &0.8370 &0.8916 &\bf 0.9046 &0.8961\\
         NP\_REC&0.7007 &0.6534 &0.7675 &\bf 0.7624 &0.7524 \\
         NP\_PRE &0.3084 &0.3579 &0.5087 &\bf 0.5589 &0.5267 \\
         NP\_F1&0.4283 &0.4625 &0.6118 &\bf 0.6450 &0.6196 \\
         P\_REC &0.7829 &0.8380 &0.8976 &\bf 0.9169 &0.9066 \\
         P\_PRE &0.9498 &0.9460 &0.9654 &\bf 0.9654 &0.9636 \\
         P\_F1 &0.8583 &0.8887 &0.9303 &\bf 0.9405 &0.9342 \\
         \hline
    \end{tabular}
    \label{tab:ablative_var_sequence}
\end{table}

\subsection{Ablative Study of Varying Segment Length}\label{segment_length}
In this subsection, we studied the relationship between sound segment length and classification performance. For this, we considered 4 different segments length (2, 4, 6, and 8 seconds) based on fixed-size input sequence of 24 and evaluated on the bowel sound dataset. The evaluation results, which are obtained after Laplace HSMM, are presented in Table \ref{tab:ablative_var_segment}. 
The experimental result shows that 8 s segments impart marginal improvement over 6 s segment for some performance metrics such as overall accuracy ( 0.9036 vs 0.8981). Nevertheless, 8 s segment length results in sharp decline in performance for NP class compared to 6 s segment length. Thus, throughout the whole work, we prefer 6 s segment length, which yields the balanced performance for both classes (P and NP) without compromising overall accuracy. From this result, we suspect that the lower-sized segments less than 6 s and higher-sized segments greater than 6 s are unable to efficiently capture the more semantic information representing the bowel sound.

\begin{table}[]
    \centering
    \scriptsize
         \caption{Comparative study of 4 different segment lengths (2, 4, 6, and 8) with our method.}
    \begin{tabular}{c|c|c|c|c}
    \hline
    \hline
         Metrics&2&4&6 &8\\
         \hline
         Acc.& 0.7972 &0.8199 & 0.8981&\bf 0.9036\\
         AUC&0.7557  &0.7615 &\bf 0.8396&0.8220 \\
         MA\_F1&0.7069  &0.7073 &\bf 0.7928&0.7695   \\
         WT\_F1&0.8127  &0.8355 & 0.9046&\bf 0.9110\\
         NP\_REC&0.6914  &0.6788 &\bf 0.7624&0.7206  \\
         NP\_PRE&0.4486  &0.4291 &\bf 0.5589&0.5048 \\
         NP\_F1&0.5442  &0.5258 &\bf 0.6450&0.5937 \\
         P\_REC&0.8197  &0.8443 & 0.9169&\bf 0.9235 \\
         P\_PRE&0.9260  &0.9384 & 0.9654&\bf 0.9683 \\
         P\_F1&0.8696 &0.8889 & 0.9405&\bf 0.9453\\
         \hline
    \end{tabular}
    \label{tab:ablative_var_segment}
\end{table}

\subsection{Ablative Study of Laplace HSMM Refinement} \label{subsec:lhsmm}
The Laplace HSMM refinement strategy is one of the major contributions of this work. We have implemented the following experiments to provide more insights into the strategy. 

\subsubsection{Comparison with Conventional HSMM Practice}
HMM, HSMM, and their extensions are widely used to improve time series classification \cite{esmael2012improving,jiang2019robust}. In conventional HMM and HSMM practices, an observation is normally referred to a predicted class label and the emission probability matrix is learned from training results. In this work, we define an observation as the probability of the segment belonging to class P, and we propose to use the Laplace distribution to model the emission probability. 

We experimentally compared the performances of HSMM with conventional and Laplace-based emission matrix. Specifically, the conventional emission matrix $E$ is obtained by
\begin{equation}
\color{black}
	E_{i j}=\frac{C_{j i} \times 100}{\sum_{i=1}^{n} C_{j i}}, \quad i, j \in \{P, NP\}
\end{equation} 
where $C$ denotes the confusion matrix of the training data. 
Table \ref{tab:different hsmm} summarizes the improvements of CNN after applying the two different types of HSMMs. With the conventional HSMM, the overall accuracy, AUC score, macro F1, and weighted F1 of CNN increased by 
0.0061, -0.0011, 0.0089, and 0.0086, respectively. By contrast, with the Laplace-based HSMM, the same metrics increased by 0.0497, 0.043, 0.0728, 0.0446,
respectively. The significant gaps prove the superiority of the Laplace-based emission matrix. 

The improvement introduced by the conventional HSMM is very limited. One explanation is that the emission probabilities, learned from the training confusion matrix, are highly concentrated because the model training performance is usually good. That means, a hidden NP or P state usually has an over 80\% chance to produce a corresponding NP or P observation. There will be a great penalty if HSMM assign a hidden state to an observation with different label. In fact, this is a dilemma. If the training stops too early, the classification results from CNN will be less reliable and the refinement will be meaningless. If the training stops too late, the emission probabilities will be concentrated. It is very difficult to control the training performance to an optimal level. 

\begin{table} [!t]
	\centering
	\scriptsize
	\caption{Comparison of Improvements by HSMM with conventional and Laplace-based emission matrix}
	\label{tab:different hsmm}
	
	\begin{tabular}{p{1.34cm}p{1.34cm}p{1.34cm}p{1.34cm}p{1.34cm}}
		\hline
		\hline
		\specialrule{0em}{1pt}{2pt}
		HSMM & ACC Improvement & AUC Improvement & MA\_F1 \quad Improvement & WT\_F1 \quad Improvement \\
		\specialrule{0em}{1pt}{1pt}
		\hline
		\specialrule{0em}{1pt}{2pt}
Conventional & + 0.0061 & - 0.0011 & + 0.0089 & + 0.0086\\
		\hline
		\specialrule{0em}{1pt}{1pt}
    	Laplace & + \textbf{0.0497}  & + \textbf{0.0438}   & + \textbf{0.0728}  & + \textbf{0.0446}   \\
		\specialrule{0em}{1pt}{1pt}
		\specialrule{0em}{1pt}{1pt}
		\hline
	\end{tabular}
\end{table}
\subsubsection{Ablative Study of Laplace HSMM}
An ablative analysis was performed to quantify the improvement introduced by the proposed Laplace HSMM strategy. We compared the CNN classification performances with and without HSMM refinement. The results are summarized in Table \ref{tab:ablative}. 
It can be observed that, after refinement, all 10 performance metrics of CNN got improved. The improvements ranged from 0.54\% to 12.89\%, where the most significant ones were made on the precision rate of non-peristalsis samples and the recall rate of peristalsis samples. This means a large portion of misclassified peristalsis samples were corrected.
The results strongly suggest that the proposed Laplace HSMM strategy has introduced comprehensive improvements to CNN. 
It helps CNN to achieve a more accurate and more reliable automatic neonatal bowel sound classification. 
 


\begin{table*} [!t]
	\centering
	\scriptsize
	\caption{Ablative Analysis of Laplace HSMM with CNN}
	\label{tab:ablative}

	\begin{tabular}{p{3.0cm}p{0.9cm}p{0.9cm}p{0.9cm}p{0.9cm}p{0.1cm}p{0.9cm}p{0.9cm}p{0.9cm}p{0.1cm}p{0.9cm}p{0.9cm}p{0.9cm}}
		\hline
		\hline
		\specialrule{0em}{1pt}{2pt}
		\multirow{2}{*}{Method} & \multirow{2}{*}{ACC} & \multirow{2}{*}{AUC} & \multirow{2}{*}{MA\_F1} & \multirow{2}{*}{WT\_F1} && \multicolumn{3}{l}{Non-peristalsis} & & \multicolumn{3}{l}{Peristalsis} \\
		\specialrule{0em}{1pt}{1pt}
		\cline{7-9}
		\cline{11-13}
		\specialrule{0em}{1pt}{2pt}
		&&&&&& REC & PRE & F1 && REC & PRE & F1 \\
		\specialrule{0em}{1pt}{1pt}
		\hline
		\specialrule{0em}{1pt}{2pt}
		CNN & 0.8484 & 0.7958 &0.7200 &0.8600 &&0.7300 & 0.4300 &0.5400 &&0.8700 &0.9600 &0.9100 \\
		\hline
		\specialrule{0em}{1pt}{1pt}
		\textbf{CNN + Laplace HSMM} & \textbf{0.8981} $\uparrow$ & \textbf{0.8396} $\uparrow$  & \textbf{0.7928} $\uparrow$  & \textbf{0.9046} $\uparrow$  && \textbf{0.7624} $\uparrow$ & \textbf{0.5589} $\uparrow$  & \textbf{0.6450} $\uparrow$  && \textbf{0.9169} $\uparrow$  & \textbf{0.9654} $\uparrow$  & \textbf{0.9405} $\uparrow$  \\
		\specialrule{0em}{1pt}{1pt}
		\specialrule{0em}{1pt}{1pt}
		\hline
	\end{tabular}
\end{table*}

\subsubsection{The Impact of Laplace Parameter}
\begin{figure}[tb]
	\centering
	\includegraphics[width=\columnwidth,height=100mm,keepaspectratio]{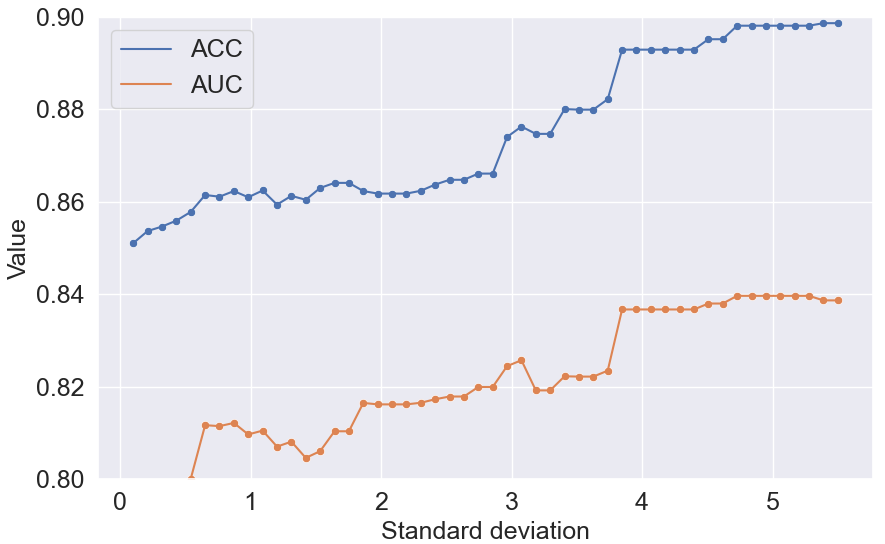}
	\caption{
		The impact of standard deviation on Laplace-based HSMM. The ACC and AUC values are from our CNN model after refinement.
	} 
	\label{fig:hsmm parameters}
\end{figure}

The Laplace distribution is a parametric function. In this subsection, we investigated the variations of the proposed HSMM performance with the changes of the standard deviation value $\sigma$ of the Laplace distribution. The result is visualized in Fig. \ref{fig:hsmm parameters}. Although there are some fluctuations, general increasing trends can be observed for both ACC and AUC scores when $\sigma$ increases from 0.1 to 5. 
Afterward, the model performance maintains at a relative stable level. The reason behind is that a small $\sigma$ leads to concentrated emission probabilities, thus limiting the refinement. With the increase of $\sigma$, the Laplace distribution becomes more spread and various hidden state combinations are allowed to be examined.

\subsubsection{Generalization Performance}
\begin{figure*}[htbp]
	\centering
	\includegraphics[width=1\textwidth]{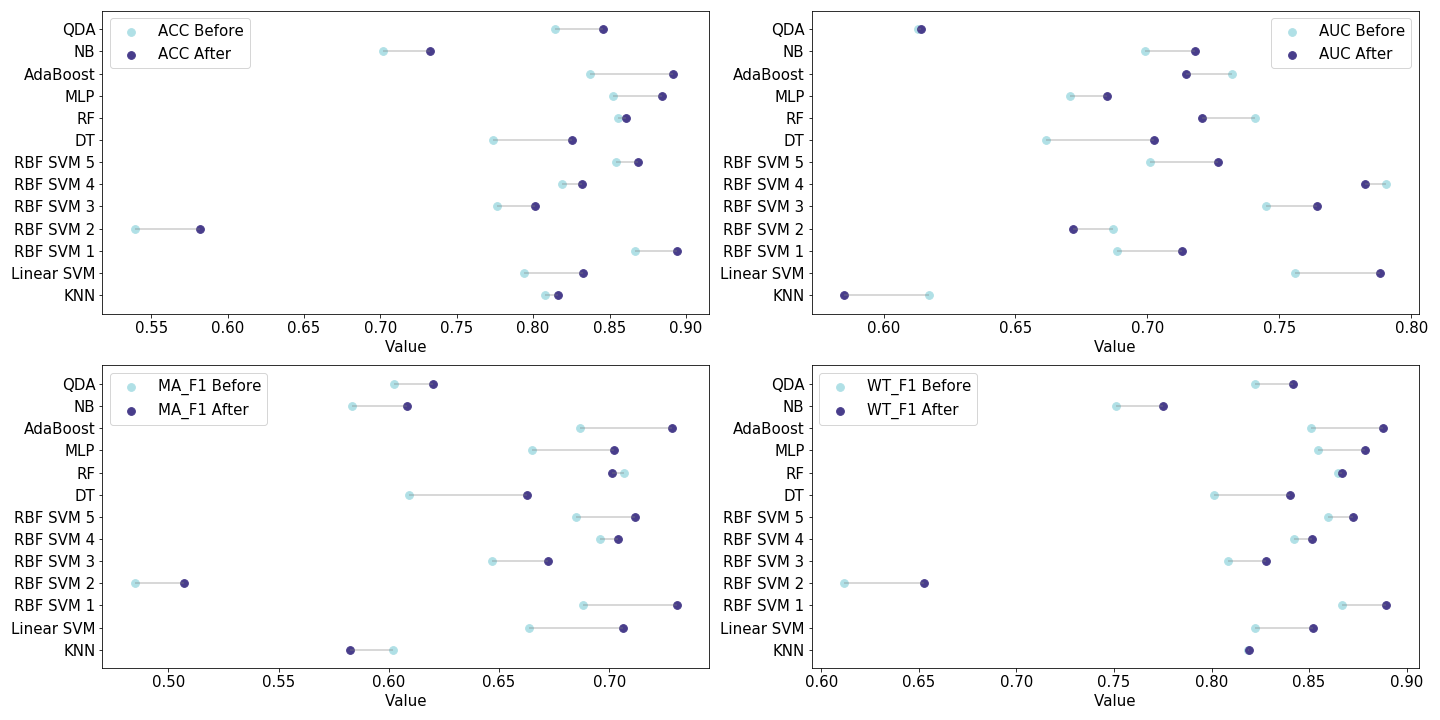}
	\caption{
		General improvements introduced by the proposed Laplace HSMM strategy.
	} 
	\label{fig:hsmm refinement}
\end{figure*}

Although the Laplace HSMM is initially proposed to improve the performance of CNN model, it can be applied to other bowel sound classification algorithms for result refinement. In this part, we investigated the generalization performance of the proposed Laplace HSMM strategy by applying it to the baselines. The results are summarized in Fig. \ref{fig:hsmm refinement}. It is worth to note that, both the ACC and WT\_F1 scores of all 13 baselines gained different levels of increases after refinement by the proposed Laplace HSMM. The most significant one is AdaBoost, with the ACC score rising from 0.8371 to 0.8913 and the WT\_F1 score rising from 0.8511 to 0.8877. Out of the total 13 baselines, 8 baselines gained improvements on the AUC score, and 11 baselines gained improvements on the MA\_F1 score, respectively. KNN and RF are the only two baselines not being improved on both AUC and MA\_F1 scores. 
The reason can be found in Table \ref{tab:results}. The two baselines performed poorly on identifying non-peristalsis samples. This might confuse the HSMM model which might treat the predicted non-peristalsis samples as noises and then label them as peristalsis. 

Despite the occasional cases, the overall result still provides the strong evidences that the proposed Laplace HSMM is able to introduce general improvements to help deliver a better and more reliable neonatal bowel sound classification.
 
\section{Discussion} \label{sec:discussion}
This paper proposed a methodology based on CNN and Laplace HSMM to detect bowel sound on infants for the first time. The experimental results have proven the effectiveness of the proposed method and demonstrated its superiority as compared to the baselines. Furthermore, we have also shown that the proposed Laplace HSMM refinement strategy can be flexibly applied to existing methods without changing their original structures to help provide a better bowel sound classification performance. 

While these results are promising, there is still room for improvements. 
Firstly, the proposed method learns to distinguish peristalsis from non-peristalsis from MFCC only. Although the effectiveness of MFCC has been proven, whether it is the best choice is still unknown because it does not contain much time domain and subband-specific information which might also present differences between peristalsis and non-peristalsis samples. Therefore, it is worth to investigate the advantages of other potential features. In fact, an investigation of the features listed in Table \ref{tab:features} has been performed. For each round of LOPOCV, feature selection is performed on training data to find out the top 20 most representative features. Although the best feature list exhibits some inter-subject variations, we can still summarize the frequenters, as shown in Table \ref{tab:ranked features}. It is clear that MFCC, Skewness, Kurtosis, and Spectral Subband Centroids and Energies are commonly determined as representative features. The finding has provided us a good direction to work on. However, to design a practical model that can make the best of each of these features is still challenging, which will be our future work.

\begin{table} [!t]
	\centering
	\scriptsize
	\caption{Frequently selected features.}
	\label{tab:ranked features}
	\begin{threeparttable}
		\begin{tabular}{p{2.8cm}p{0.7cm}|p{2.8cm}p{0.7cm}}
			\hline
			\hline
			Feature & Count & Feature & Count  \\
			\hline
			MFCC\_8 & 49 & SSC\_1511.0Hz - 1743.0Hz & 31 \\
			MFCC\_9 & 49 & MFCC\_10 & 31 \\
			MFCC\_11 & 49 & SSC\_398.0Hz - 513.0Hz & 30 \\
			MFCC\_12 & 49 & Kurtosis & 27 \\
			MFCC\_18 & 49 & SSC\_1743.0Hz - 2000.0Hz & 19 \\
			SSC\_74.0Hz - 155.0Hz\tnote{1} & 49 & SSC\_344.0Hz - 454.0Hz & 17 \\
			SSC\_199.0Hz - 293.0Hz & 49 & SSC\_36.0Hz - 113.0Hz & 14 \\
			SSC\_454.0Hz - 575.0Hz  & 49 & MFCC\_2 & 9 \\
			SSC\_641.0Hz - 782.0Hz & 49 & MFCC\_14 & 5 \\
			SB\_EN\_199.0Hz - 293.0Hz\tnote{2} & 47 & MFCC\_13 & 3 \\
			SSC\_710.0Hz - 858.0Hz & 47 & SB\_EN\_155.0Hz - 245.0Hz & 2 \\
			SSC\_0.0Hz - 74.0Hz & 46 & SSC\_513.0Hz - 641.0Hz & 2 \\
			Skewness & 44 & SSC\_113.0Hz - 199.0Hz & 2 \\
			SSC\_245.0Hz - 344.0Hz & 43 & MFCC\_6 & 1 \\
			SSC\_293.0Hz - 398.0Hz & 40 & MFCC\_7 & 1 \\
			SSC\_575.0Hz - 710.0Hz & 39 & MFCC\_21 & 1 \\
			SSC\_858.0Hz - 1022.0Hz & 37 & SSC\_1624.0Hz - 1868.0Hz & 1 \\
			\hline
		\end{tabular}
		\begin{tablenotes}
			\footnotesize
			\item[1] SSC denotes the spectral subband centroid.
			\item[2] SB\_EN denotes the spectral subband energy.
		\end{tablenotes}
	\end{threeparttable}
\end{table}

Secondly, the experimental abdominal sounds were collected in NICU with a commercial DS without noise cancellation function. They are very noisy, containing rub noise, alarm noise, monitor noise, and talk noise. This may be one possible reason of why most of the baselines failed to give a good performance even if the used features have been proven effective in adult bowel sound detection. 
We did not specially handle the noise problem because the neonatal abdominal sound is very weak. The sound recording might lose important information while the noise is suppressed. 
Although the proposed method has been proven to be able to achieve a promising result on these noisy recordings, having good quality data is still necessary. This will allow our model to allocate more learning capacities to catch the key differences between peristalsis and non-peristalsis samples instead of wasting resources on reducing the influence of noise.
Moreover, the presence of noises can also impose a great challenge for the follow-up analysis, in which we need to further determine whether the peristalsis sound is disease related or not. Therefore, in our next study, we will use a more advanced DS and denoising methods to reduce the noise interference. 

Thirdly, the abdominal sound segments are highly imbalanced, with nearly 88\% being peristalsis samples. This explains why the proposed method reported a precision rate of just 0.5589 for non-peristalsis samples, given that the recall rates for both classes were quite good. Even though just a small portion of peristalsis samples were misclassified, there was still a large impact to the precision of non-peristalsis samples. 
Moreover, the imbalance problem limited the available data because class balancing by downsampling had to be performed to reduce model training bias. The model complexity is thus limited. 
Our team of neonatologists are currently working to solve the problem by continuously collecting data. Although this will be a very time-consuming process, it will bring a lot of benefits, not only to our current work but also to our future study of pathology of bowel sound. 

Fourthly, the proposed method was developed only based on the abdominal sounds recorded from the right lower quadrant. Whether the abdominal sounds from the other three quadrants (right upper, left upper and left lower) help to deliver a more accurate peristalsis detection and how to develop a mode to effectively analyze the abdominal sounds from all four quadrants are still to be explored. 
 
This work has shown how automatic analysis of neonatal bowel sound can be done from DS collected abdominal sounds.  DS auscultation will have been a mainstream methodology for diagnosis of bowel condition in the foreseen future. Although point-of-care ultrasound can provide more detailed information on bowel peristalsis, bowel wall thickness and bowel vascularity \cite{priyadarshi2019neonatologist}, it can be only performed by medical professionals within hospital settings. By contrast, DS auscultation is much more convenient without imposing constraints on professional knowledge and geographical locations. It facilitates the development and application of telehealth. One can use DS to record bowel sound in home environment and send out the recording with the help of smartphones and high-speed networks for remote auscultation. 
The proposed bowel sound detection model is one of the major contributions of this study, which can greatly improve the efficiency of the bowel condition tele-diagnosis process. 
The current work is a preliminary step towards automated classification of different types of peristalsis sounds and pathological bowel conditions.  
With this model, clinicians can directly listen to their sounds of interest, instead of going over every detail. The ability to use other biological sounds such as heart and lung sounds  \cite{grooby2020neonatal} with the proposed methodologies could be explored.

\textcolor{black}{Automated detection of peristalsis sounds is an important step towards automated detection of neonatal bowel conditions from abdominal sounds. The next steps will be to differentiate and characterise various types of bowel sounds, validate the results against other modalities such as ultrasound and develop diagnostic models to detect bowel dysfunctions, toward developing a clinical decision support system.}

\section{Conclusion} \label{sec:conclusion}
This paper presented an effective neonatal bowel sound detection method, which uses CNNs
to perform an initial classification which is then refined with Laplace HSMM. The proposed deep learning model enables automatic learning of useful patterns 
to distinguish peristalsis from non-peristalsis.
The Laplace HSMM refinement strategy optimizes the predicted events by considering their hidden time dependency. Abdominal sounds recorded from 49 newborn infants admitted to our tertiary NICU were used as experimental data. The results showed that the proposed method can accurately identify peristalsis and non-peristalsis sounds with an overall accuracy of around 90\% and an AUC score of around 84\%. In addition, the Laplace HSMM refinement strategy has been proven to be able to introduce general improvements to other bowel sound detection models to help perform a more reliable detection. 

This work opens opportunities for further research to develop continuous acoustic monitoring or ``mapping" to quantitate patterns of bowel acoustic signatures that may allow pre-clinical detection of NEC and Septicaemia.


\bibliographystyle{IEEEtran}
\bibliography{bibliography}

\begin{thebibliography}{10}
\providecommand{\url}[1]{#1}
\csname url@samestyle\endcsname
\providecommand{\newblock}{\relax}
\providecommand{\bibinfo}[2]{#2}
\providecommand{\BIBentrySTDinterwordspacing}{\spaceskip=0pt\relax}
\providecommand{\BIBentryALTinterwordstretchfactor}{4}
\providecommand{\BIBentryALTinterwordspacing}{\spaceskip=\fontdimen2\font plus
\BIBentryALTinterwordstretchfactor\fontdimen3\font minus
  \fontdimen4\font\relax}
\providecommand{\BIBforeignlanguage}[2]{{%
\expandafter\ifx\csname l@#1\endcsname\relax
\typeout{** WARNING: IEEEtran.bst: No hyphenation pattern has been}%
\typeout{** loaded for the language `#1'. Using the pattern for}%
\typeout{** the default language instead.}%
\else
\language=\csname l@#1\endcsname
\fi
#2}}
\providecommand{\BIBdecl}{\relax}
\BIBdecl

\bibitem{liu2018bowel}
J.~Liu, Y.~Yin, H.~Jiang, H.~Kan, Z.~Zhang, P.~Chen, B.~Zhu, and Z.~Wang,
  ``Bowel sound detection based on mfcc feature and lstm neural network,'' in
  \emph{2018 IEEE Biomedical Circuits and Systems Conference (BioCAS)}.\hskip
  1em plus 0.5em minus 0.4em\relax IEEE, 2018, pp. 1--4.

\bibitem{queensland2019routine}
\BIBentryALTinterwordspacing
Q.~C. Guidelines. (2019) Routine newborn assessment. [Online]. Available:
  \url{https://www.health.qld.gov.au/__data/assets/pdf_file/0029/141689/g-newexam.pdf}
\BIBentrySTDinterwordspacing

\bibitem{dumas2013feasibility}
J.~DUMAS, K.~M. HILL, R.~S. ADREZIN, J.~ALBA, R.~CURRY, E.~CAMPAGNA,
  C.~FERNANDES, V.~LAMBA, and L.~EISENFELD, ``Feasibility of an electronic
  stethoscope system for monitoring neonatal bowel sounds.'' \emph{Connecticut
  Medicine}, vol.~77, no.~8, 2013.

\bibitem{hill2008stethoscope}
J.~M. Hill, A.~Maloney, K.~Stephens, R.~Adrezin, and L.~Eisenfeld,
  ``Stethoscope for monitoring neonatal abdominal sounds,'' in \emph{Proc
  IAJC-IJME Int Conf}, vol.~9, no.~1, 2008, pp. 5--11.

\bibitem{calvert2020necrotising}
W.~Calvert, K.~Sampat, M.~Jones, C.~Baillie, G.~Lamont, and P.~Losty,
  ``Necrotising enterocolitis--a 15-year outcome report from a uk specialist
  centre,'' \emph{Acta Paediatrica}, 2020.

\bibitem{priyadarshi2020continuous}
A.~Priyadarshi, M.~Hinder, N.~Badawi, M.~Luig, and M.~Tracy, ``Continuous
  positive airway pressure belly syndrome: Challenges of a changing paradigm,''
  \emph{International Journal of Clinical Pediatrics}, vol.~9, no.~1, pp.
  9--15, 2020.

\bibitem{ramanathan2019digital}
A.~Ramanathan, L.~Zhou, F.~Marzbanrad, R.~Roseby, K.~Tan, A.~Kevat, and
  A.~Malhotra, ``Digital stethoscopes in paediatric medicine,'' \emph{Acta
  Paediatrica}, vol. 108, no.~5, pp. 814--822, 2019.

\bibitem{yin2018bowel}
Y.~Yin, H.~Jiang, S.~Feng, J.~Liu, P.~Chen, B.~Zhu, and Z.~Wang, ``Bowel sound
  recognition using svm classification in a wearable health monitoring
  system.'' \emph{Sci. China Inf. Sci.}, vol.~61, no.~8, pp. 084\,301--1, 2018.

\bibitem{dimoulas2008bowel}
C.~Dimoulas, G.~Kalliris, G.~Papanikolaou, V.~Petridis, and A.~Kalampakas,
  ``Bowel-sound pattern analysis using wavelets and neural networks with
  application to long-term, unsupervised, gastrointestinal motility
  monitoring,'' \emph{Expert Systems with Applications}, vol.~34, no.~1, pp.
  26--41, 2008.

\bibitem{ulusar2013real}
U.~D. Ulusar, M.~Canpolat, M.~Yaprak, S.~Kazanir, and G.~Ogunc, ``Real-time
  monitoring for recovery of gastrointestinal tract motility detection after
  abdominal surgery,'' in \emph{2013 7th International Conference on
  Application of Information and Communication Technologies}.\hskip 1em plus
  0.5em minus 0.4em\relax IEEE, 2013, pp. 1--4.

\bibitem{ulusar2014recovery}
U.~D. Ulusar, ``Recovery of gastrointestinal tract motility detection using
  naive bayesian and minimum statistics,'' \emph{Computers in biology and
  medicine}, vol.~51, pp. 223--228, 2014.

\bibitem{sheu2014higher}
M.-J. Sheu, P.-Y. Lin, J.-Y. Chen, C.-C. Lee, and B.-S. Lin,
  ``Higher-order-statistics-based fractal dimension for noisy bowel sound
  detection,'' \emph{IEEE Signal Processing Letters}, vol.~22, no.~7, pp.
  789--793, 2014.

\bibitem{wainberg2018deep}
M.~Wainberg, D.~Merico, A.~Delong, and B.~J. Frey, ``Deep learning in
  biomedicine,'' \emph{Nature biotechnology}, vol.~36, no.~9, pp. 829--838,
  2018.

\bibitem{sitaula2021attention}
C.~Sitaula and M.~B. Hossain, ``Attention-based vgg-16 model for covid-19 chest
  x-ray image classification,'' \emph{Applied Intelligence}, vol.~51, no.~5,
  pp. 2850--2863, 2021.

\bibitem{han2020deep}
X.~Han, Y.~Hu, L.~Foschini, L.~Chinitz, L.~Jankelson, and R.~Ranganath, ``Deep
  learning models for electrocardiograms are susceptible to adversarial
  attack,'' \emph{Nature Medicine}, pp. 1--4, 2020.

\bibitem{hochreiter1997long}
S.~Hochreiter and J.~Schmidhuber, ``Long short-term memory,'' \emph{Neural
  computation}, vol.~9, no.~8, pp. 1735--1780, 1997.

\bibitem{lahav2015questionable}
A.~Lahav, ``Questionable sound exposure outside of the womb: frequency analysis
  of environmental noise in the neonatal intensive care unit,'' \emph{Acta
  paediatrica}, vol. 104, no.~1, pp. e14--e19, 2015.

\bibitem{3M20183M}
\BIBentryALTinterwordspacing
3M. (2018) {3M\textsuperscript{\texttrademark}
  Littmann\textsuperscript{\textregistered} Electronic Stethoscope Model 3200.
  St. Paul, MN, USA: 3M}. [Online]. Available:
  \url{https://www.littmann.com/3M/en_US/littmann-stethoscopes/}
\BIBentrySTDinterwordspacing

\bibitem{nijmegen2020elan}
\BIBentryALTinterwordspacing
T.~L.~A. Nijmegen: Max Planck Institute~for Psycholinguistics. (2020) Elan
  (version 5.9) [computer software]. [Online]. Available:
  \url{https://archive.mpi.nl/tla/elan}
\BIBentrySTDinterwordspacing

\bibitem{ittichaichareon2012speech}
C.~Ittichaichareon, S.~Suksri, and T.~Yingthawornsuk, ``Speech recognition
  using mfcc,'' in \emph{International Conference on Computer Graphics,
  Simulation and Modeling}, 2012, pp. 135--138.

\bibitem{lecun1995convolutional}
Y.~LeCun, Y.~Bengio \emph{et~al.}, ``Convolutional networks for images, speech,
  and time series,'' \emph{The handbook of brain theory and neural networks},
  vol. 3361, no.~10, p. 1995, 1995.

\bibitem{allwood2018advances}
G.~Allwood, X.~Du, K.~M. Webberley, A.~Osseiran, and B.~J. Marshall, ``Advances
  in acoustic signal processing techniques for enhanced bowel sound analysis,''
  \emph{IEEE reviews in biomedical engineering}, vol.~12, pp. 240--253, 2018.

\bibitem{anders2020comparison}
F.~Anders, M.~Hlawitschka, and M.~Fuchs, ``Comparison of artificial neural
  network types for infant vocalization classification,'' \emph{IEEE/ACM
  Transactions on Audio, Speech, and Language Processing}, vol.~29, pp. 54--67,
  2020.

\bibitem{goodfellow2016deep}
I.~Goodfellow, Y.~Bengio, A.~Courville, and Y.~Bengio, \emph{Deep
  learning}.\hskip 1em plus 0.5em minus 0.4em\relax MIT press Cambridge, 2016,
  vol.~1.

\bibitem{rabiner1989tutorial}
L.~R. Rabiner, ``A tutorial on hidden markov models and selected applications
  in speech recognition,'' \emph{Proceedings of the IEEE}, vol.~77, no.~2, pp.
  257--286, 1989.

\bibitem{yu2010hidden}
S.-Z. Yu, ``Hidden semi-markov models,'' \emph{Artificial intelligence}, vol.
  174, no.~2, pp. 215--243, 2010.

\bibitem{forney2005viterbi}
G.~D. Forney~Jr, ``The viterbi algorithm: A personal history,'' \emph{arXiv
  preprint cs/0504020}, 2005.

\bibitem{ppgb2011jitter}
\BIBentryALTinterwordspacing
ppgb. (2011) Definitions of jitter measurements. [Online]. Available:
  \url{http://145.100.59.186/praat/manual/Voice_2__Jitter.html}
\BIBentrySTDinterwordspacing

\bibitem{ppgb2003shimmer}
\BIBentryALTinterwordspacing
------. (2003) Definitions of shimmer measurements. [Online]. Available:
  \url{http://145.100.59.186/praat/manual/Voice_3__Shimmer.html}
\BIBentrySTDinterwordspacing

\bibitem{kim2011estimation}
K.~S. Kim, J.~H. Seo, S.~H. Ryu, M.~H. Kim, and C.~G. Song, ``Estimation
  algorithm of the bowel motility based on regression analysis of the jitter
  and shimmer of bowel sounds,'' \emph{Computer methods and programs in
  biomedicine}, vol. 104, no.~3, pp. 426--434, 2011.

\bibitem{kim2011non}
K.-S. Kim, J.-H. Seo, and C.-G. Song, ``Non-invasive algorithm for bowel
  motility estimation using a back-propagation neural network model of bowel
  sounds,'' \emph{Biomedical engineering online}, vol.~10, no.~1, p.~69, 2011.

\bibitem{kim2012awareness}
K.-S. Kim, H.-J. Park, H.~S. Kang, and C.-G. Song, ``Awareness system for bowel
  motility estimation based on artificial neural network of bowel sounds,'' in
  \emph{4th International Conference on Awareness Science and
  Technology}.\hskip 1em plus 0.5em minus 0.4em\relax IEEE, 2012, pp. 185--188.

\bibitem{yin2015bowel}
Y.~Yin, W.~Yang, H.~Jiang, and Z.~Wang, ``Bowel sound based digestion state
  recognition using artificial neural network,'' in \emph{2015 IEEE Biomedical
  Circuits and Systems Conference (BioCAS)}.\hskip 1em plus 0.5em minus
  0.4em\relax IEEE, 2015, pp. 1--4.

\bibitem{lin2013enhancing}
B.-S. Lin, M.-J. Sheu, C.-C. Chuang, K.-C. Tseng, and J.-Y. Chen, ``Enhancing
  bowel sounds by using a higher order statistics-based radial basis function
  network,'' \emph{IEEE Journal of Biomedical and Health Informatics}, vol.~17,
  no.~3, pp. 675--680, 2013.

\bibitem{dimoulas2007long}
C.~Dimoulas, G.~Kalliris, G.~Papanikolaou, and A.~Kalampakas, ``Long-term
  signal detection, segmentation and summarization using wavelets and fractal
  dimension: A bioacoustics application in gastrointestinal-motility
  monitoring,'' \emph{Computers in Biology and Medicine}, vol.~37, no.~4, pp.
  438--462, 2007.

\bibitem{ranta2001wavelet}
R.~Ranta, C.~Heinrich, V.~Louis-Dorr, D.~Wolf, and F.~Guillemin,
  ``Wavelet-based bowel sounds denoising, segmentation and characterization,''
  in \emph{2001 Conference Proceedings of the 23rd Annual International
  Conference of the IEEE Engineering in Medicine and Biology Society},
  vol.~2.\hskip 1em plus 0.5em minus 0.4em\relax IEEE, 2001, pp. 1903--1906.

\bibitem{will2020a}
\BIBentryALTinterwordspacing
W.~Koehrsen. (2020) A feature selection tool for machine learning in python.
  [Online]. Available:
  \url{https://towardsdatascience.com/a-feature-selection-tool-for-machine-learning-in-python-b64dd23710f0}
\BIBentrySTDinterwordspacing

\bibitem{fawcett2006introduction}
T.~Fawcett, ``An introduction to roc analysis,'' \emph{Pattern recognition
  letters}, vol.~27, no.~8, pp. 861--874, 2006.

\bibitem{demvsar2006statistical}
J.~Dem{\v{s}}ar, ``Statistical comparisons of classifiers over multiple data
  sets,'' \emph{The Journal of Machine Learning Research}, vol.~7, pp. 1--30,
  2006.

\bibitem{esmael2012improving}
B.~Esmael, A.~Arnaout, R.~K. Fruhwirth, and G.~Thonhauser, ``Improving time
  series classification using hidden markov models,'' in \emph{2012 12th
  International Conference on Hybrid Intelligent Systems (HIS)}.\hskip 1em plus
  0.5em minus 0.4em\relax IEEE, 2012, pp. 502--507.

\bibitem{jiang2019robust}
D.~Jiang, Y.-n. Lu, M.~Yu, and W.~Yuanyuan, ``Robust sleep stage classification
  with single-channel eeg signals using multimodal decomposition and hmm-based
  refinement,'' \emph{Expert Systems with Applications}, vol. 121, pp.
  188--203, 2019.

\bibitem{priyadarshi2019neonatologist}
A.~Priyadarshi, S.~Rogerson, M.~Hinder, and M.~Tracy, ``Neonatologist performed
  point-of-care bowel ultrasound: Is the time right?'' \emph{Australasian
  Journal of Ultrasound in Medicine}, vol.~22, no.~1, pp. 15--25, 2019.

\bibitem{grooby2020neonatal}
E.~Grooby, J.~He, J.~Kiewsky, D.~Fattahi, L.~Zhou, A.~King, A.~Ramanathan,
  A.~Malhotra, G.~A. Dumont, and F.~Marzbanrad, ``Neonatal heart and lung sound
  quality assessment for robust heart and breathing rate estimation for
  telehealth applications,'' \emph{IEEE Journal of Biomedical and Health
  Informatics}, 2020.

\end{thebibliography}

\end{document}